%% file: LvTsSub.tex
\documentclass[aps,prd,twocolumn,nofootinbib,floatfix]{revtex4-2}
\usepackage{graphicx}
\usepackage{amsmath}
\usepackage[colorlinks=true]{hyperref}
\usepackage{placeins}

\begin{document}
\title{Ionisation potentials and energy levels of ions of heavy and superheavy elements Te, I, Po, At, Lv and Ts.}
\author{G. K. Vong}\email{g.vong@unsw.edu.au}
\author{V. A. Dzuba}\email{v.dzuba@unsw.edu.au}
\author{V. V. Flambaum}\email{v.flambaum@unsw.edu.au}
\affiliation{School of Physics, University of New South Wales, Sydney 2052, Australia}

\begin{abstract}
We calculate the energy levels and successive ionisation potentials (IPs) of ions of the three heaviest known Group 16 and 17 elements using a theoretical approach that combines the linearised coupled-cluster method, configuration interaction, and perturbation theory. Our calculations address critical gaps in the available data on the electronic structure of the superheavy elements livermorium (Lv) and tennessine (Ts), as well as their lighter homologues polonium (Po) and astatine (At). To assess the accuracy of our methods, we perform analogous calculations for the lighter homologues tellurium (Te) and iodine (I), for which both experimental and reliable theoretical data are available for comparison.

\end{abstract}

\maketitle

\section{Introduction}
Superheavy elements (SHEs), with atomic numbers $Z>104$, are of considerable interest in both atomic and nuclear physics. Their large nuclear charges 
 and large number of electrons give rise to strong electron correlations coupled with extreme relativistic effects, which may lead to exotic atomic properties that deviate from trends observed in lighter elements~\cite{smits2023,smits2024,ackermann2024,ye2025}. Furthermore, SHEs also provide a means to probe the predicted "island of stability", a region around $Z=114$  and $N=184$ ($N$ is neutron number) where nuclear shell models suggest the existence of longer-lived isotopes~\cite{ackermann2024,DFW17-IS}. 

Although elements up to $Z=118$ have been synthesised, there is no experimental spectroscopic data for SHEs with $Z>102$. This is primarily due to their extremely short half-lives and low production rates, which make direct measurements of electronic structure highly challenging. As a result, until substantial experimental advancements occur, theoretical studies remain the primary tool for investigating SHE properties~\cite{smits2023,smits2024}.

Reliable data on the properties of SHEs ions are especially scarce. Compared to neutral SHEs, studying their ions is equally important, and in some respects even more so. Ionisation achieved by removing one or more electrons, reduces the screening of the nuclear charge, alters electron correlations and modifies relativistic effects such as spin-orbit couplings. 
The energy spectrum and ionisation potentials (IPs) of an atom are fundamental to understanding its electronic structure. Calculating energy levels and IPs across multiple charge states provides a stringent test for relativistic many-body methods in superheavy atoms. Moreover, IPs are important for predicting a wide range of properties relevant to nuclear physics, atomic physics and chemistry, such as chemical reactivity.

In this work, we extend our previous calculations from Ref.~\cite{LvTsNeutral} for the neutral Group 16 and 17 elements to their ions, addressing critical gaps in the available data. Specifically, we calculate energy levels and successive ionisation potentials by removing valence  celectrons one at a time until reaching configurations with a single valence electron. This covers ions with charges from +1 to +6 for the Group 16 elements (Te, Po, Lv) and from +1 to +7 for the Group 17 elements (Te, Po, Lv). Our calculations employ a combination of the linearised coupled-cluster single-double (SD) method~\cite{Saf-CC,Dzu-CI-SD14} with configuration interaction perturbation theory (CIPT)~\cite{CIPT}. This SD+CIPT method balances accuracy and computational efficiency, as demonstrated in our earlier study of the neutral atoms. To validate the accuracy of our approach, we use calculations for the lightest homologues considered, Te and I, as benchmarks, where more experimental and reliable theoretical data are available.

\section{Computational Method}
Our computational approach is largely similar to that used for the neutral atoms, with a few key differences for ions. The neutral atom configurations of the outermost electron is $ns^2np^4$ for Te, Po, and Lv, and $ns^2np^5$ for I, At, and Ts, with $n=5$ for Te and I, $n=6$ for Po and At, and $n=7$ for Lv and Ts. 

For neutral atoms, the absence of excitations from the closed $ns$ subshell in the NIST~\cite{NIST_ASD} indicates that such states lie very high in the spectrum, allowing us to effectively treat the $ns$ subshell as part of the core. In contrast, for ions, spectra of Te II and I II in the NIST database show low-lying states involving excitations from the $ns$ subshell. Therefore, in our ionic calculations, the active valence space includes both the closed $s$-shell and the open $p$-shell.

The method involves solving the linearised coupled-cluster single-double (SD) equations first for the core then for valence states, followed by employing the configuration interaction (CI) technique to treat the valence electrons. Perturbation theory (PT) is used to incorporate contributions from high-energy valence states perturbatively. The core potential, RHF potential, SD equations, correlation operators $\hat{\Sigma}_1$ and $\hat{\Sigma}_2$, and 
single-electron basis states remain the same for all ions of a specific element. 
The primary difference lies in the number of electrons included in the final CI calculation. This is achieved by defining the appropriate reference configurations for each ion and generating a full set of basis configurations with the same parity and number of electrons. 
{This is done by exciting one or two electrons from reference configurations into single-electron basis states and then constructing for each configuration all possible single-determinant  many-electron basis states of definite total angular momentum $J$ and parity.}
The CI calculations for each ion are done by successively removing valence electrons until the valence space is depleted.

For completeness, we summarise the key aspects of the method below; for further details, refer to our previous work on neutral atoms~\cite{LvTsNeutral}. Our calculations begin with the relativistic Hartree-Fock (RHF) procedure for the closed-shell core, which is the same for all ions of a given element. The RHF Hamiltonian is given by
\begin{equation}
    \hat{H}^{\mathrm{RHF}} =  c \boldsymbol{\alpha} \cdot \hat{\mathbf{p}} + (\beta - 1) mc^2 + V_{\text{nuc}} + V_{\text{core}},
    \label{h1}
\end{equation}
where $V_{\text{nuc}}$ is the nuclear potential derived from integrating the nuclear charge density described by a Fermi distribution, $V^{\text{core}}$ is the self-consistent RHF potential of the closed-shell core after all valence electrons are removed. This corresponds to the so-called $V^{N-M}$ approximation~\cite{Dzu05}, where $N$ is the total number of electrons and $M$ the number of valence electrons. 
$M$ varies from one to five for Te, Po and Lv ions, and from one to six for I, At and Ts ions.


{Although the  $V^{N-M}$  approximation starts from a less precise description of electron wave functions -  an apparent drawback when many valence electrons are present - its great strength is the straightforward, reliable treatment of core-valence correlations. In practice, this benefit outweighs the loss of initial accuracy.

Inside the core region, the electrostatic potential created by the valence electrons is nearly constant; a constant potential leaves the core- electron wave functions essentially unchanged. Consequently, the main source of error in the $V^{N-M}$ scheme is not an imperfect core description but the finite quality of the valence-electron basis. That deficiency can be suppressed simply by employing a sufficiently large and flexible basis set.

By contrast, in the standard $V^{N}$ approach - where all electrons are included self-consistently - one must contend with large "subtraction" diagrams that slow the convergence of the many-body expansion. Because such diagrams are absent in the $V^{N-M}$  formulation, calculations converge faster and more stably. A full discussion of these points is provided in Ref.~\cite{Dzu05}.}


Single-electron basis states are calculated in the frozen-core potential using the standard B-spline technique~\cite{B-splines}. We construct these states as linear combinations of B-splines by diagonalising the RHF Hamiltonian matrix in the B-spline basis. To ensure the basis is sufficiently saturated, we compute 40 B-spline states of order 9 within a spherical box of radius 40$a_B$ (where $a_B$ is the Bohr radius) for each partial wave up to $l_\text{max} = 6$. These basis states serve as the foundation for both the SD and CI calculations.

The many-electron wavefunction is expanded using the SD method:
\begin{equation}
\begin{aligned}
|\Psi \rangle &= \left[ 1 + \sum_{ma} \rho_{ma} a_{m}^{\dagger} a_{a} + \frac{1}{2} \sum_{mnab} \rho_{mnab} a_{m}^{\dagger} a_{n}^{\dagger} a_{b} a_{a} \right. \\
& \quad \left. + \sum_{m \neq \nu} \rho_{m \nu} a_{m}^{\dagger} a_{\nu} + \sum_{mnb} \rho_{mnvb} a_{m}^{\dagger} a_{n}^{\dagger} a_{b} a_{\nu} \right] |\Phi_0\rangle,
\end{aligned}
\end{equation}
where $|\Phi_0\rangle$ is the reference RHF state, indices $a$, $b$ refer to core states, indices $m$, $n$, $k$, $l$ refer to excited states and index $v$ refers to the valence state. The expansion coefficients $\rho$ correspond to the single and double electron excitations from the core and from the valence state. The coefficients are found by solving the SD equations~\cite{Dzu-CI-SD14}, which also give the one- and two- electron operators $\hat{\Sigma}_1$ and $\hat{\Sigma}_2$, respectively. 
{The purpose of the SD calculations is to determine these operators. The $\hat{\Sigma}$ operators describe core-valence correlations. Specifically, the $\hat{\Sigma}_1$ operator accounts for the correlation interaction between a particular valence electron and the core. The $\hat{\Sigma}_2$ operator represents the screening of the Coulomb interaction between two valence electrons by the core electrons.} 


Including these correlation operators, the CI Hamiltonian becomes
\begin{equation}
    \hat{H}^{\mathrm{CI}} = \sum_i^{M} \left(\hat{H}^{\mathrm{RHF}}+ \hat{\Sigma}_1 \right)_i + \sum_{i<j}^{M} \left( \frac{e^2}{|\boldsymbol{r}_i-\boldsymbol{r}_j|} + \hat{\Sigma}_{2ij} \right),
    \label{HCI}
\end{equation}
where $i$ and $j$ are the indices for valence electrons. 

{Solving full-scale CI equations for a large number of valence electrons (up to six in present work) is a challenging task.}
To perform calculations efficiently, we use the CIPT method~\cite{CIPT}, in which neglecting the off-diagonal matrix elements between high-energy states allows one to reduce the size of the effective CI matrix by many orders of magnitude. Meanwhile, the contribution of the high-energy states are included perturbatively. 
The many-electron wavefunction is expanded over low-energy states $P$ and over high-energy states $Q$:
\begin{equation}
|\Psi_m\rangle = \sum_{i \in P} c_{im} |\phi_i\rangle + \sum_{j \in Q} c_{jm} |\phi_j\rangle,
\end{equation}
where $\phi_i$ are single-determinant many-electron basis functions constructed from the provided reference configuration (or set of reference configurations) by exciting one or two electrons.

By neglecting the off-diagonal matrix elements from states in $Q$ the matrix is reduced to 
\begin{equation}
H_{ij}^\text{eff} = H_{ij} + \sum_{k \in Q} \frac{H_{ik} H_{kj}}{E - E_k},
\label{e:e2}
\end{equation}
where $H_{ij}$ is the CI matrix for states in $P$, $H_{ik}$ are the matrix elements between states in $P$ and $Q$, $E_k = \langle \phi_k | \hat{H}^\mathrm{CI} | \phi_k \rangle$ and $E$ is energy of the state of interest. 
{The relative size of $P$ and $Q$ subspaces comes from a compromise between accuracy and efficiency. For one, two or three valence electrons $P$ could cover all basis space, no need for $Q$. The natural choice for large number of valence electrons is $P \ll Q$.}

The expansion coefficients $c_{im}$,  and energies $E_m$ in Eq.~\eqref{e:e2} are determined by solving the eigenvalue problem 
\begin{equation}
    (\hat{H}^{\text{eff}} - E I)X=0,
    \label{eigeneq}
\end{equation}
where $X$ is the vector of expansion coefficients $c_{im}$. Note that the $\hat{H}^{\text{eff}}$ matrix (\ref{e:e2}) depends on unknown energy $E$.
Therefore, the equations (\ref{e:e2},\ref{eigeneq}) should be solved iteratively until convergence in $E$ is achieved.

Breit interactions and quantum electrodynamic (QED) radiative corrections are included using a radiative potential to account for magnetic interactions, retardation, vacuum polarization, and self-energy effects (see Refs.~\cite{radpot,DzuEtAl01a,DzuFlaSaf06,Etotal} for details). Although these corrections contribute to the excitation energies less than the uncertainties from correlations for Lv and Ts, they are included to ensure completeness.

For each level, Land\'{e} $g$-factors are calculated and compared to both experimental values and the non-relativistic expression
\begin{equation}
g = 1 + \frac{J(J+1)-L(L+1)+S(S+1)}{2J(J+1)}.
\label{e:gf}
\end{equation}
The Land\'{e} $g$-factor is obtained as the expectation value of the magnetic dipole interaction operator (M1): \mbox{$\hat{H}_{M1}=\boldsymbol{\mu}\cdot \mathbf{B}$}, where $\boldsymbol{\mu}$ is the atomic magnetic moment and $\mathbf{B}$ is the external magnetic field. For a valence electron state $v$, with total angular momentum $J$, the relativistic $g$-factor is given by

\begin{equation}
    g_J = \frac{1}{\sqrt{J(2J+1)(J+1)}}  \langle v||\hat{H}_{M1}||v\rangle /B.
    \label{landeg}
\end{equation}
Since $L$ and $S$ are not good quantum numbers in relativistic calculations, we find values of $L$ and $S$ that make the non-relativistic expression \eqref{e:gf} best match the calculated $g_J$ in Eq.~\eqref{landeg}. This helps in assigning term symbols and grouping levels into fine structure multiplets. Further details can be found in Ref.~\cite{LvTsNeutral}. 

In the final step, ionisation potentials are obtained as the difference between the ground-state energies of successive ions, from +1 to +6 for Lv and from +1 to +7 for Ts.

\section{Ionisation Potentials}

The calculated successive IPs are presented in \mbox{Table~\ref{IP}.} The calculation of the IPs involves simply changing the number of electrons in the CI calculations and taking the difference between the ground state energies of each successive ion. Thus, agreement of calculated IPs with experiment are also an indication of the accuracy of the spectra calculations. The first ionisation potentials were calculated in our previous paper \cite{LvTsNeutral}.

Relativistic energy shifts scale as $(Z\alpha)^2$, making them significantly more pronounced in the SHEs Lv and Ts compared to their lighter homologues. For the atoms considered in this work, the relative magnitude of relativistic effects increases from approximately 0.14 in Te to 0.73 in Ts. In particular, a substantial spin-orbit splitting ocurrs within the $7p$ shell, where the $7p_{1/2}$ subshell is lowered in energy while the $7p_{3/2}$ subshell is raised. This effect is evident in the calculated ionisation potentials: for Lv, the large increase between $\text{IP}_2$ and $\text{IP}_3$ corresponds with the removal of a $7p_{1/2}$ electron after removing two $7p_{3/2}$ electrons. Ts also exhibits a similar jump between $\text{IP}_3$ and $\text{IP}_4$. Note that while this feature is also present in the IPs of Po and At, the relative differences are smaller due to weaker relativistic effects. In the lighter elements Te and I, this effect is barely discernible.

Overall, we find the relative difference between theory and experiment to be $\sim 3$\%. Our calculated values for Lv and Ts are on the same level as the most recent advanced calculations. It is natural to assume the same level of accuracy for the IPs of Lv and Ts. 

\section{Energy levels of Te and I ions}

Te and I are the heaviest atoms with similar electron structures to Lv and Ts that have experimental data for their ions. As such, comparing calculated energy levels to experimental values provides a good indicator of accuracy. The results for the ions of Te and I are presented in Tables \ref{Te II} to  \ref{I VII}.

Comparison with experiment shows the average relative difference between theory and experiment over all ions is $\sim1.44$\% of the excitation energy for tellurium, similarly for iodine it is $\sim 1.36$\%. We should expect similar accuracy for heavier ions of Po, At, Lv, and Ts.

\section{Energy levels of Po and At ions}

Experimental data for the ions of Po and At is non-existent, so the calculated energy levels (presented in Tables \ref{Po II} to  \ref{At VII}) serve to provide accurate theoretical values. Compared to the the lighter homologues of Te and I, relativistic effects start to play a role because they are proportional to $(Z\alpha)^2$.  These effects shift energies and cause configuration mixing, particularly in high energy states. Coupled with the lack of experimental $g$-factors, this makes highly excited states difficult to identify. However, since accuracy from the lighter homologues was very good and our study dedicated to the neutral atoms \cite{LvTsNeutral} showed good agreement with the existing data, we expect good accuracy for our calculated values.

\section{Energy levels of Lv and Ts ions}

 The calculated spectra of Lv and Ts ions are presented in Tables \ref{Lv II} to  \ref{Ts VII}. As there are no available experimental or theoretical data for these spectra, our results provide the first predictions for the electronic structure of these ions. We observe a continuation of the trend found in the neutral atoms, where the energy spacing between different configurations becomes increasingly compressed in the heavier homologues. The computed $g$-factors differ from the non-relativistic predictions, indicating strong configuration mixing and deviation from the LS-coupling scheme.

\section{Conclusion}
In this work, we have presented calculations of the energy levels and successive ionisation potentials for ions of the heaviest known Group 16 and 17 elements, including Te, I, Po, At, Lv, and Ts.
Using a combination of the linearised coupled-cluster method with configuration interaction and perturbation theory (SD+CIPT), we extended our previous analysis of neutral atoms to multiple ionised states, up to +6 for Group 16 and +7 for Group 17 elements. To validate the accuracy of our approach, we benchmarked our calculations against Te and I ions, where comparison with experimental data showed an average deviation of approximately 1.4\%. For the heavier homologues Po and At, and the superheavy elements Lv and Ts, our calculations fill critical gaps in the data and provide accurate theoretical predictions for their electronic spectra and ionisation potentials. Our results reveal strong relativistic effects, most notably very large spin-orbit splittings in the $7p$ shell of Lv and Ts, which manifest as significant shifts in their energy levels and ionisation potentials. 

\begin{acknowledgments}
This work was supported by the Australian Research Council Grant No. DP230101058.
\end{acknowledgments}

\bibliographystyle{apsrev4-2}
\input{LvTsSub.bbl}


\appendix
\begin{table*}[ht]
    \caption{Ionisation potentials (in cm$^{-1}$) of I, A and Ts ions. Entries marked by a * denote a value that was determined by interpolation, extrapolation or by semi-empirical calculation.}
    \label{IP}
\begin{tabular}{ccccl}
    \hline \hline
    Ionisation Stage & Atom & Present & Expt. & Other Calculations \\ \hline
    I & Te & 70 147 & 72 669 &  \\
     & I & 72 925 & 84 295  &  \\
     & Po & 67 901 & 67 896  &  \\
     & At & 71 254 & 75 151  &  \\
     & Lv & 54 416 &  & 47 601 \cite{Hangele}, 55 290 \cite{Borschevsky_2015}, 53 470 \cite{Dyall2012}, 59 600 \cite{Z116}, 60 000 \cite{Fricke1976} \\
     & Ts & 58 986 &  & 61 730 \cite{Borschevsky_2015}, 59 520 \cite{Dyall2012} , 61 310 \cite{Liu_2006}, 61 600 \cite{Z117}, 61 100 \cite{Mitin}, 61 700 \cite{Pyper}, 62 000 \cite{Fricke1976} \\
     &  &  &  &  \\
\hline \hline
    \end{tabular}
\end{table*}

\begin{table}[htpb]
    \caption{Te II}
    \label{Te II}
    \centering
    \begin{tabular}{cccccccc}
    \hline \hline
&  &  & \multicolumn{2}{c}{Energy ($\text{cm}^{-1}$)} &  & \multicolumn{2}{c}{$g_J$} \\
\cline{4-5} \cline{7-8}
No. & Conf.             & Term           & Calc.    & NIST     &  & Calc.  & NIST   \\ \hline
1   & $5s^{2}5p^{3}$    & $^4$S$^{\rm o}_{3/2}$  & 0        & 0        &  & 1.9348 & 1.93   \\
2   & $5s^{2}5p^{3}$    & $^2$D$^{\rm o}_{3/2}$  & 10 550   & 10 222   &  & 0.9308 &        \\
3   & $5s^{2}5p^{3}$    & $^2$D$^{\rm o}_{5/2}$  & 11 782   & 12 422   &  & 1.2000 &        \\
4   & $5s^{2}5p^{3}$    & $^2$P$^{\rm o}_{1/2}$  & 20 308   & 20 547   &  & 0.6667 &        \\
5   & $5s^{2}5p^{3}$    & $^2$P$^{\rm o}_{3/2}$  & 24 656   & 24 032    &  & 1.2677 & 1.27   \\
6   & $5s5p^{4}$        & $^4$P$_{5/2}$  & 71 503   & 71 193   &  & 1.5841 & 1.59   \\
\hline \hline
    \end{tabular}
\end{table}

\begin{table}[hb]
    \caption{Te III}
    \label{Te III}
    \centering
    \begin{tabular}{cccccccc}
    \hline \hline
&  &  & \multicolumn{2}{c}{Energy ($\text{cm}^{-1}$)} &  & \multicolumn{2}{c}{$g_J$} \\
\cline{4-5} \cline{7-8}
No. & Conf.             & Term           & Calc.    & NIST     &  & Calc.  & NIST   \\ \hline
1   & $5s^{2}5p^{2}$    & $^1$S$_{0}$    & 0        & 0        &  & 0.0000 &        \\
2   & $5s^{2}5p^{2}$    & $^3$P$_{1}$    & 4 872    & 4 751    &  & 1.5000 &        \\
3   & $5s^{2}5p^{2}$    & $^3$P$_{2}$    & 8 364    & 8 165    &  & 1.3840 &        \\
4   & $5s^{2}5p^{2}$    & $^3$D$_{2}$    & 17 881   & 17 358   &  & 1.1160 &        \\
5   & $5s^{2}5p^{2}$    & $^1$S$_{0}$    & 31 071   &   &  & 0.0000 &        \\
6   & $5s5p^{3}$        & $^5$S$^{\rm o}_{2}$    & 65 608   &          &  & 1.9831 &        \\
\hline \hline
    \end{tabular}
\end{table}

\begin{table}[htpb]
\    \caption{Te IV}
    \label{Te IV}
    \centering
    \begin{tabular}{cccccccc}
    \hline \hline
&  &  & \multicolumn{2}{c}{Energy ($\text{cm}^{-1}$)} &  & \multicolumn{2}{c}{$g_J$} \\
\cline{4-5} \cline{7-8}
No. & Conf.             & Term           & Calc.    & NIST     &  & Calc.  & NIST   \\ \hline
1   & $5s^{2}5p$        & $^2$P$^{\rm o}_{1/2}$  & 0        & 0        &  & 0.6667 &        \\
2   & $5s^{2}5p$        & $^2$P$^{\rm o}_{3/2}$  & 9 536    & 9 223    &  & 1.3333 &        \\
3   & $5s5p^{2}$        & $^4$P$_{1/2}$  & 70 152   &          &  & 2.6412 &        \\
4   & $5s5p^{2}$        & $^4$P$_{3/2}$  & 74 493   &          &  & 1.7258 &        \\
5   & $5s5p^{2}$        & $^4$P$_{5/2}$  & 79 156   &          &  & 1.5698 &        \\
6   & $5s5p^{2}$        & $^2$D$_{3/2}$  & 93 372   & 92 772   &  & 0.8189 &        \\
\hline \hline
    \end{tabular}
\end{table}

\begin{table}[htpb]
    \caption{Te V}
    \label{Te V}
    \centering
    \begin{tabular}{ccccccc}
    \hline \hline
&  &  & \multicolumn{2}{c}{Energy ($\text{cm}^{-1}$)} &  & $g_J$ \\
\cline{4-5} \cline{7-7}
No. & Conf.             & Term           & Calc.    & NIST     &  & Calc.  \\ \hline
1   & $5s^{2}$          & $^1$S$_{0}$    & 0        & 0        &  & 0.0000 \\
2   & $5s5p$            & $^3$P$^{\rm o}_{0}$    & 75 549   & 75 111   &  & 0.0000 \\
3   & $5s5p$            & $^3$P$^{\rm o}_{1}$    & 78 580   & 78 025    &  & 1.4877 \\
4   & $5s5p$            & $^3$P$^{\rm o}_{2}$    & 86 748   & 86 004     &  & 1.5000 \\
5   & $5s5p$            & $^1$P$^{\rm o}_{1}$    & 110 786  & 111 707  &  & 1.0123 \\
6   & $5p^{2}$          & $^1$S$_{0}$    & 176 039  & 176 255  &  & 0.0000 \\
7   & $5p^{2}$          & $^3$P$_{1}$    & 182 303  & 182 421  &  & 1.5000 \\
8   & $5p^{2}$          & $^3$D$_{2}$    & 183 378  & 182 804  &  & 1.1826 \\
9   & $5p^{2}$          & $^3$D$_{2}$    & 193 063  & 192 597  &  & 1.3160 \\
10  & $5p^{2}$          & $^1$S$_{0}$    & 215 034  & 215 061   &  & 0.0000 \\
11  & $5s5d$            & $^3$D$_{1}$    & 215 973  & 215 611  &  & 0.5000 \\
12  & $5s5d$            & $^3$D$_{2}$    & 216 251  & 216 137  &  & 1.1664 \\
13  & $5s5d$            & $^3$D$_{3}$    & 217 353  & 216 992  &  & 1.3333 \\
14  & $5s5d$            & $^1$D$_{2}$    & 231 243  & 232 054   &  & 1.0017 \\
15  & $5s6s$            & $^3$S$_{1}$    & 241 370  & 240 839  &  & 2.0000 \\
16  & $5s6s$            & $^1$S$_{0}$    & 247 392  & 246 751  &  & 0.0000 \\
17  & $5s4f$            & $^3$F$^{\rm o}_{2}$    & 258 913  &          &  & 0.6667 \\
18  & $5s4f$            & $^3$F$^{\rm o}_{3}$    & 258 946  &          &  & 1.0832 \\
19  & $5s4f$            & $^3$F$^{\rm o}_{4}$    & 258 955  &          &  & 1.2500 \\
20  & $5s4f$            & $^1$F$^{\rm o}_{3}$    & 262 673  &          &  & 1.0002 \\
\hline \hline
    \end{tabular}
\end{table}

\begin{table}[htpb]
    \caption{Te VI}
    \label{Te VI}
    \centering
    \begin{tabular}{ccccccc}
    \hline \hline
&  &  & \multicolumn{2}{c}{Energy ($\text{cm}^{-1}$)} &  & $g_J$ \\
\cline{4-5} \cline{7-7}
No. & Conf.             & Term           & Calc.    & NIST     &  & Calc.  \\ \hline
1   & $5s$              & $^2$S$_{1/2}$  & 0        & 0        &  & 2.0000 \\
2   & $5p$              & $^2$P$^{\rm o}_{1/2}$  & 93 537   & 93 336   &  & 0.6667 \\
3   & $5p$              & $^2$P$^{\rm o}_{3/2}$  & 105 575  & 105 150  &  & 1.3333 \\
4   & $5d$              & $^2$D$_{3/2}$  & 238 780  & 238 081   &  & 0.8000 \\
5   & $5d$              & $^2$D$_{5/2}$  & 240 441  & 239 725  &  & 1.2000 \\
6   & $4f$              & $^2$F$^{\rm o}_{5/2}$  & 267 116  &          &  & 0.8571 \\
7   & $4f$              & $^2$F$^{\rm o}_{7/2}$  & 267 116  &          &  & 1.1429 \\
8   & $6s$              & $^2$S$_{1/2}$  & 280 055  & 278 439  &  & 2.0000 \\
9   & $6p$              & $^2$P$^{\rm o}_{1/2}$  & 315 653  & 314 190  &  & 0.6667 \\
10  & $6p$              & $^2$P$^{\rm o}_{3/2}$  & 320 120  & 318 589  &  & 1.3333 \\
11  & $6d$              & $^2$D$_{3/2}$  & 375 337  &          &  & 0.8000 \\
12  & $6d$              & $^2$D$_{5/2}$  & 376 073  &          &  & 1.2000 \\
13  & $5f$              & $^2$F$^{\rm o}_{5/2}$  & 377 490  &          &  & 0.8571 \\
14  & $5f$              & $^2$F$^{\rm o}_{7/2}$  & 377 616  &          &  & 1.1429 \\
15  & $7s$              & $^2$S$_{1/2}$  & 392 127  &          &  & 2.0000 \\
16  & $7p$              & $^2$P$^{\rm o}_{1/2}$  & 409 744  &          &  & 0.6667 \\
17  & $5g$              & $^2$G$_{9/2}$  & 411 006  &          &  & 1.1111 \\
18  & $5g$              & $^2$G$_{7/2}$  & 411 009  &          &  & 0.8889 \\
19  & $7p$              & $^2$P$^{\rm o}_{3/2}$  & 412 038  &          &  & 1.3333 \\
20  & $6f$              & $^2$F$^{\rm o}_{5/2}$  & 439 828  &          &  & 0.8571 \\
\hline \hline
    \end{tabular}
\end{table}

\newpage

\begin{table}[htpb]
    \caption{I II}
    \label{I II}
    \centering
    \begin{tabular}{cccccccc}
    \hline \hline
&  &  & \multicolumn{2}{c}{Energy ($\text{cm}^{-1}$)} &  & \multicolumn{2}{c}{$g_J$} \\
\cline{4-5} \cline{7-8}
No. & Conf.             & Term           & Calc.    & NIST     &  & Calc.  & NIST   \\ \hline
1   & $5s^{2}5p^{4}$    & $^3$P$_{2}$    & 0        & 0        &  & 1.4476 & 1.457  \\
2   & $5s^{2}5p^{4}$    & $^1$S$_{0}$    & 6 540    & 6 448    &  & 0.0000 &        \\
3   & $5s^{2}5p^{4}$    & $^3$P$_{1}$    & 7 440    & 7 087     &  & 1.5000 & 1.51   \\
4   & $5s^{2}5p^{4}$    & $^1$D$_{2}$    & 14 554   & 13 727   &  & 1.0524 & 1.046  \\
5   & $5s^{2}5p^{4}$    & $^1$S$_{0}$    & 30 854   & 29 501   &  & 0.0000 &        \\
6   & $5s5p^{5}$        & $^3$P$^{\rm o}_{2}$    & 83 530   & 81 908   &  & 1.5392 & 1.54   \\
\hline \hline
    \end{tabular}
\end{table}

\begin{table}[htpb]
    \caption{I III}
    \label{I III}
    \centering
    \begin{tabular}{ccccccc}
    \hline \hline
&  &  & \multicolumn{2}{c}{Energy ($\text{cm}^{-1}$)} &  & $g_J$ \\
\cline{4-5} \cline{7-7}
No. & Conf.             & Term           & Calc.    & NIST     &  & Calc.  \\ \hline
1   & $5s^{2}5p^{3}$    & $^6$D$^{\rm o}_{3/2}$  & 0        & 0        &  & 1.9034 \\
2   & $5s^{2}5p^{3}$    & $^6$F$^{\rm o}_{3/2}$  & 12 192   & 11 711   &  & 0.9694 \\
3   & $5s^{2}5p^{3}$    & $^2$D$^{\rm o}_{5/2}$  & 14 345   & 14 901   &  & 1.2000 \\
4   & $5s^{2}5p^{3}$    & $^2$P$^{\rm o}_{1/2}$  & 24 218   & 24 299   &  & 0.6667 \\
5   & $5s^{2}5p^{3}$    & $^4$D$^{\rm o}_{3/2}$  & 30 467   & 29 637   &  & 1.2605 \\
6   & $5s5p^{4}$        & $^4$P$_{5/2}$  & 86 138   & 85 555   &  & 1.5812 \\
7   & $5s5p^{4}$        & $^4$P$_{3/2}$  & 91 482   & 90 964   &  & 1.7036 \\
8   & $5s5p^{4}$        & $^4$P$_{1/2}$  & 93 381   & 92 902   &  & 2.6337 \\
9   & $5s5p^{4}$        & $^2$D$_{3/2}$  & 103 614  & 103 470  &  & 0.9237 \\
10  & $5s5p^{4}$        & $^2$D$_{5/2}$  & 106 842  & 106 619  &  & 1.2085 \\
11  & $5s^{2}5p^{2}5d$  & $^4$F$_{3/2}$  & 109 855  & 110 130  &  & 0.5606 \\
12  & $5s^{2}5p^{2}5d$  & $^4$D$_{3/2}$  & 110 581  & 111 365  &  & 1.1385 \\
13  & $5s^{2}5p^{2}5d$  & $^4$F$_{5/2}$  & 111 248  & 111 806  &  & 1.0986 \\
14  & $5s^{2}5p^{2}5d$  & $^2$P$_{1/2}$  & 112 473  & 113 322  &  & 0.6682 \\
15  & $5s^{2}5p^{2}5d$  & $^4$F$_{7/2}$  & 114 206  & 115 432  &  & 1.2568 \\
16  & $5s^{2}5p^{2}5d$  & $^4$F$_{5/2}$  & 115 269  & 116 130  &  & 0.9484 \\
17  & $5s^{2}6s5p^{2}$  & $^2$S$_{1/2}$  & 116 568  & 116 557  &  & 1.9863 \\
18  & $5s^{2}5p^{2}5d$  & $^4$F$_{9/2}$  & 117 352  &          &  & 1.3101 \\
19  & $5s^{2}5p^{2}5d$  & $^4$F$_{7/2}$  & 117 455  & 118 621  &  & 1.2535 \\
20  & $5s^{2}5p^{2}5d$  & $^2$P$_{1/2}$  & 118 538  & 119 649  &  & 0.5401 \\
\hline \hline
    \end{tabular}
\end{table}

\begin{table}[htpb]
    \caption{I IV}
    \label{I IV}
    \centering
    \begin{tabular}{ccccccc}
    \hline \hline
&  &  & \multicolumn{2}{c}{Energy ($\text{cm}^{-1}$)} &  & $g_J$ \\
\cline{4-5} \cline{7-7}
No. & Conf.             & Term           & Calc.    & NIST     &  & Calc.  \\ \hline
1   & $5s^{2}5p^{2}$    & $^1$S$_{0}$    & 0        & 0        &  & 0.0000 \\
2   & $5s^{2}5p^{2}$    & $^3$P$_{1}$    & 6 946    & 6 828    &  & 1.5000 \\
3   & $5s^{2}5p^{2}$    & $^3$P$_{2}$    & 11 234   & 10 982   &  & 1.3579 \\
4   & $5s^{2}5p^{2}$    & $^3$D$_{2}$    & 23 181   & 22 532   &  & 1.1421 \\
5   & $5s^{2}5p^{2}$    & $^1$S$_{0}$    & 38 048   & 37 177   &  & 0.0000 \\
6   & $5s5p^{3}$        & $^5$S$^{\rm o}_{2}$    & 78 892   & 78 084    &  & 1.9756 \\
7   & $5s5p^{3}$        & $^3$D$^{\rm o}_{1}$    & 99 670   & 99 047    &  & 0.6051 \\
8   & $5s5p^{3}$        & $^3$D$^{\rm o}_{2}$    & 100 191  & 99 542   &  & 1.2185 \\
9   & $5s5p^{3}$        & $^3$D$^{\rm o}_{3}$    & 103 083  & 102 387  &  & 1.3329 \\
10  & $5s5p^{3}$        & $^1$S$^{\rm o}_{0}$    & 115 236  & 114 658  &  & 0.0000 \\
11  & $5s5p^{3}$        & $^3$D$^{\rm o}_{2}$    & 115 481  & 115 013   &  & 1.2638 \\
12  & $5s5p^{3}$        & $^3$P$^{\rm o}_{1}$    & 116 078  & 115 478  &  & 1.4227 \\
13  & $5s5p^{3}$        & $^3$D$^{\rm o}_{2}$    & 123 418  & 123 187  &  & 1.1741 \\
14  & $5s^{2}5p5d$      & $^3$F$^{\rm o}_{3}$    & 133 356  & 133 737  &  & 1.0902 \\
15  & $5s5p^{3}$        & $^3$P$^{\rm o}_{1}$    & 134 310  & 135 677  &  & 1.7053 \\
16  & $5s^{2}5p5d$      & $^3$F$^{\rm o}_{4}$    & 139 344  &          &  & 1.2500 \\
17  & $5s^{2}5p5d$      & $^1$P$^{\rm o}_{1}$    & 142 744  & 143 392  &  & 0.9456 \\
18  & $5s^{2}6s5p$      & $^1$S$^{\rm o}_{0}$    & 148 969  & 149 049   &  & 0.0000 \\
19  & $5s^{2}5p5d$      & $^3$D$^{\rm o}_{3}$    & 150 299  & 151 872  &  & 1.3009 \\
20  & $5s^{2}5p5d$      & $^1$S$^{\rm o}_{0}$    & 151 414  & 153 038   &  & 0.0000 \\
\hline \hline
    \end{tabular}
\end{table}

\begin{table}[htpb]
    \caption{I V}
    \label{I V}
    \centering
    \begin{tabular}{ccccccc}
    \hline \hline
&  &  & \multicolumn{2}{c}{Energy ($\text{cm}^{-1}$)} &  & $g_J$ \\
\cline{4-5} \cline{7-7}
No. & Conf.             & Term           & Calc.    & NIST     &  & Calc.  \\ \hline
1   & $5s^{2}5p$        & $^2$P$^{\rm o}_{1/2}$  & 0        & 0        &  & 0.6667 \\
2   & $5s^{2}5p$        & $^2$P$^{\rm o}_{3/2}$  & 12 534   & 12 222   &  & 1.3333 \\
3   & $5s5p^{2}$        & $^4$P$_{1/2}$  & 81 616   & 81 018    &  & 2.6317 \\
4   & $5s5p^{2}$        & $^4$P$_{3/2}$  & 87 594   & 85 556   &  & 1.7241 \\
5   & $5s5p^{2}$        & $^4$P$_{5/2}$  & 93 378   & 92 558   &  & 1.5600 \\
6   & $5s5p^{2}$        & $^2$D$_{3/2}$  & 109 362  & 108 780  &  & 0.8257 \\
7   & $5s5p^{2}$        & $^2$D$_{5/2}$  & 112 626  & 111 831  &  & 1.2396 \\
8   & $5s5p^{2}$        & $^2$P$_{1/2}$  & 124 793  & 125 704  &  & 1.1160 \\
9   & $5s5p^{2}$        & $^2$S$_{1/2}$  & 138 164  & 138 328  &  & 1.5857 \\
10  & $5s5p^{2}$        & $^2$P$_{3/2}$  & 138 185  & 139 398  &  & 1.3090 \\
11  & $5s^{2}5d$        & $^2$D$_{3/2}$  & 153 434  & 154 50   &  & 0.8079 \\
12  & $5s^{2}5d$        & $^2$D$_{5/2}$  & 154 870  & 155 462  &  & 1.2004 \\
13  & $5s^{2}6s$        & $^2$S$_{1/2}$  & 176 813  & 176 814  &  & 2.0000 \\
14  & $5s^{2}4f$        & $^2$F$^{\rm o}_{7/2}$  & 187 799  &          &  & 1.1429 \\
15  & $5s^{2}4f$        & $^2$F$^{\rm o}_{5/2}$  & 187 803  &          &  & 0.8572 \\
16  & $5p^{3}$          & $^2$P$^{\rm o}_{3/2}$  & 204 496  & 204 475  &  & 1.3776 \\
17  & $5p^{3}$          & $^2$P$^{\rm o}_{3/2}$  & 209 810  & 209 983  &  & 1.4244 \\
18  & $5p^{3}$          & $^2$D$^{\rm o}_{5/2}$  & 210 697  & 210 90   &  & 1.1987 \\
19  & $5s^{2}6p$        & $^2$P$^{\rm o}_{1/2}$  & 211 101  & 211 216  &  & 0.6668 \\
20  & $5s^{2}6p$        & $^2$P$^{\rm o}_{3/2}$  & 214 965  & 215 053   &  & 1.3298 \\
\hline \hline
    \end{tabular}
\end{table}

\begin{table}[htpb]
    \caption{I VI}
    \label{I VI}
    \centering
    \begin{tabular}{ccccccc}
    \hline \hline
&  &  & \multicolumn{2}{c}{Energy ($\text{cm}^{-1}$)} &  & $g_J$ \\
\cline{4-5} \cline{7-7}
No. & Conf.             & Term           & Calc.    & NIST     &  & Calc.  \\ \hline
1   & $5s^{2}$          & $^1$S$_{0}$    & 0        & 0        &  & 0.0000 \\
2   & $5s5p$            & $^3$P$^{\rm o}_{0}$    & 86 185   & 85 666   &  & 0.0000 \\
3   & $5s5p$            & $^3$P$^{\rm o}_{1}$    & 89 873   & 89 262   &  & 1.4840 \\
4   & $5s5p$            & $^3$P$^{\rm o}_{2}$    & 100 551  & 99 685   &  & 1.5000 \\
5   & $5s5p$            & $^1$P$^{\rm o}_{1}$    & 126 549  & 127 424  &  & 1.0159 \\
6   & $5p^{2}$          & $^1$S$_{0}$    & 199 871  & 200 085   &  & 0.0000 \\
7   & $5p^{2}$          & $^3$P$_{1}$    & 208 375  & 208 474  &  & 1.5000 \\
8   & $5p^{2}$          & $^3$D$_{2}$    & 210 058  & 209 432  &  & 1.1964 \\
9   & $5p^{2}$          & $^3$D$_{2}$    & 222 591  & 221 984  &  & 1.3024 \\
10  & $5p^{2}$          & $^1$S$_{0}$    & 245 853  & 245 658  &  & 0.0000 \\
11  & $5s5d$            & $^3$D$_{1}$    & 252 159  & 251 817  &  & 0.5000 \\
12  & $5s5d$            & $^3$D$_{2}$    & 252 686  & 252 540  &  & 1.1662 \\
13  & $5s5d$            & $^3$D$_{3}$    & 254 097  & 253 752  &  & 1.3333 \\
14  & $5s5d$            & $^1$D$_{2}$    & 269 285  & 274 162  &  & 1.0017 \\
15  & $5s4f$            & $^3$F$^{\rm o}_{2}$    & 272 252  &          &  & 0.6667 \\
16  & $5s4f$            & $^3$F$^{\rm o}_{3}$    & 272 356  &          &  & 1.0833 \\
17  & $5s4f$            & $^3$F$^{\rm o}_{4}$    & 272 407  &          &  & 1.2500 \\
18  & $5s4f$            & $^1$F$^{\rm o}_{3}$    & 278 556  &          &  & 1.0000 \\
19  & $5s6s$            & $^3$S$_{1}$    & 296 982  & 296 253  &  & 2.0000 \\
20  & $5s6s$            & $^1$S$_{0}$    & 303 712  & 305 966  &  & 0.0000 \\
\hline \hline
    \end{tabular}
\end{table}

\begin{table}[htpb]
    \caption{I VII}
    \label{I VII}
    \centering
    \begin{tabular}{ccccccc}
    \hline \hline
&  &  & \multicolumn{2}{c}{Energy ($\text{cm}^{-1}$)} &  & $g_J$ \\
\cline{4-5} \cline{7-7}
No. & Conf.             & Term           & Calc.    & NIST     &  & Calc.  \\ \hline
1   & $5s$              & $^2$S$_{1/2}$  & 0        & 0        &  & 2.0000 \\
2   & $5p$              & $^2$P$^{\rm o}_{1/2}$  & 105 145  & 104 960  &  & 0.6667 \\
3   & $5p$              & $^2$P$^{\rm o}_{3/2}$  & 120 409  & 119 958  &  & 1.3333 \\
4   & $4f$              & $^2$F$^{\rm o}_{5/2}$  & 271 604  & 271 000    &  & 0.8571 \\
5   & $4f$              & $^2$F$^{\rm o}_{7/2}$  & 271 869  & 271 200  &  & 1.1429 \\
6   & $5d$              & $^2$D$_{3/2}$  & 274 734  & 274 019   &  & 0.8000 \\
7   & $5d$              & $^2$D$_{5/2}$  & 276 988  & 276 256  &  & 1.2000 \\
8   & $6s$              & $^2$S$_{1/2}$  & 337 235  & 335 376  &  & 2.0000 \\
9   & $6p$              & $^2$P$^{\rm o}_{1/2}$  & 378 892  & 377 185  &  & 0.6667 \\
10  & $6p$              & $^2$P$^{\rm o}_{3/2}$  & 384 758  & 382 985  &  & 1.3333 \\
11  & $5f$              & $^2$F$^{\rm o}_{5/2}$  & 437 899  & 456 450  &  & 0.8571 \\
12  & $5f$              & $^2$F$^{\rm o}_{7/2}$  & 438 178  &          &  & 1.1429 \\
13  & $6d$              & $^2$D$_{3/2}$  & 450 459  & 456 450  &  & 0.8000 \\
14  & $6d$              & $^2$D$_{5/2}$  & 451 468  & 467 453  &  & 1.2000 \\
15  & $7s$              & $^2$S$_{1/2}$  & 477 115  & 476 331  &  & 2.0000 \\
16  & $5g$              & $^2$G$_{9/2}$  & 490 185  & 487 810  &  & 1.1111 \\
17  & $5g$              & $^2$G$_{7/2}$  & 490 191  & 487 800  &  & 0.8889 \\
18  & $7p$              & $^2$P$^{\rm o}_{1/2}$  & 498 021  &          &  & 0.6667 \\
19  & $7p$              & $^2$P$^{\rm o}_{3/2}$  & 501 015  & 495 769  &  & 1.3333 \\
20  & $6f$              & $^2$F$^{\rm o}_{5/2}$  & 526 848  & 495 769  &  & 0.8571 \\
\hline \hline
    \end{tabular}
\end{table}

\newpage

\begin{table}[htpb]
    \caption{Po II}
    \label{Po II}
    \centering
    \begin{tabular}{ccccc}
    \hline \hline
No. & Conf.             & Term           & Energy ($\text{cm}^{-1}$) & $g_J$    \\ \hline
1   & $6s^{2}6p^{3}$    & $^4$P$^{\rm o}_{3/2}$  & 0        & 1.5624   \\
2   & $6s^{2}6p^{3}$    & $^2$P$^{\rm o}_{3/2}$  & 17 968   & 1.3072   \\
3   & $6s^{2}6p^{3}$    & $^2$D$^{\rm o}_{5/2}$  & 22 936   & 1.2001   \\
4   & $6s^{2}6p^{3}$    & $^2$P$^{\rm o}_{1/2}$  & 30 520   & 0.6667   \\
5   & $6s^{2}6p^{3}$    & $^4$D$^{\rm o}_{3/2}$  & 49 262   & 1.2641   \\
6   & $6s^{2}7s6p^{2}$  & $^2$S$_{1/2}$  & 62 198   & 2.0849   \\
7   & $6s^{2}6p^{2}6d$  & $^2$D$_{3/2}$  & 72 794   & 0.7852   \\
8   & $6s^{2}6p^{2}6d$  & $^2$D$_{5/2}$  & 75 715   & 1.2344   \\
9   & $6s^{2}6p^{2}7p$  & $^2$P$^{\rm o}_{1/2}$  & 79 106   & 0.6730   \\
10  & $6s^{2}7s6p^{2}$  & $^4$P$_{3/2}$  & 80 848   & 1.6828   \\
11  & $6s^{2}7s6p^{2}$  & $^2$P$_{1/2}$  & 82 223   & 1.2510   \\
12  & $6s^{2}6p^{2}8p$  & $^2$P$^{\rm o}_{3/2}$  & 84 575   & 1.3393   \\
13  & $6s^{2}7s6p^{2}$  & $^4$D$_{5/2}$  & 85 469   & 1.3950   \\
14  & $6s^{2}7s6p^{2}$  & $^6$F$_{3/2}$  & 86 359   & 1.0074   \\
15  & $6s^{2}6p^{2}6d$  & $^4$D$_{5/2}$  & 88 471   & 1.4042   \\
16  & $6s^{2}6p^{2}6d$  & $^6$F$_{3/2}$  & 89 856   & 1.0306   \\
17  & $6s^{2}6p^{2}6d$  & $^4$F$_{7/2}$  & 91 295   & 1.2851   \\
18  & $6s^{2}6p^{2}6d$  & $^4$F$_{5/2}$  & 91 491   & 0.9512   \\
19  & $6s^{2}6p^{2}6d$  & $^4$D$_{1/2}$  & 91 999   & 0.3138   \\
20  & $6s^{2}6p^{2}6d$  & $^2$F$_{7/2}$  & 94 640   & 1.1706   \\
\hline \hline
    \end{tabular}
\end{table}

\begin{table}[htpb]
    \caption{Po III}
    \label{Po III}
    \centering
    \begin{tabular}{ccccc}
    \hline \hline
No. & Conf.             & Term           & Energy ($\text{cm}^{-1}$) & $g_J$    \\ \hline
1   & $6s^{2}6p^{2}$    & $^1$S$_{0}$    & 0        & 0.0000   \\
2   & $6s^{2}6p^{2}$    & $^3$P$_{1}$    & 20 441   & 1.4999   \\
3   & $6s^{2}6p^{2}$    & $^3$D$_{2}$    & 24 968   & 1.2345   \\
4   & $6s^{2}6p^{2}$    & $^3$D$_{2}$    & 49 896   & 1.2655   \\
5   & $6s^{2}6p^{2}$    & $^1$S$_{0}$    & 61 375   & 0.0000   \\
6   & $6s6p^{3}$        & $^5$P$^{\rm o}_{2}$    & 97 249   & 1.7621   \\
7   & $6s^{2}7s6p$      & $^1$S$^{\rm o}_{0}$    & 106 060  & 0.0000   \\
8   & $6s^{2}7s6p$      & $^1$P$^{\rm o}_{1}$    & 106 530  & 1.2350   \\
9   & $6s6p^{3}$        & $^1$P$^{\rm o}_{1}$    & 108 208  & 0.9026   \\
10  & $6s^{2}6p6d$      & $^3$F$^{\rm o}_{2}$    & 109 750  & 0.7856   \\
11  & $6s^{2}6p6d$      & $^3$P$^{\rm o}_{2}$    & 112 445  & 1.3759   \\
12  & $6s^{2}6p6d$      & $^3$F$^{\rm o}_{3}$    & 114 790  & 1.1617   \\
13  & $6s6p^{3}$        & $^1$P$^{\rm o}_{1}$    & 124 103  & 0.9101   \\
14  & $6s6p^{3}$        & $^3$D$^{\rm o}_{2}$    & 125 623  & 1.1893   \\
15  & $6s6p^{3}$        & $^5$F$^{\rm o}_{3}$    & 127 148  & 1.2665   \\
16  & $6s^{2}6p7p$      & $^3$D$_{1}$    & 127 716  & 0.6722   \\
17  & $6s^{2}6p7p$      & $^1$S$_{0}$    & 131 942  & 0.0000   \\
18  & $6s6p^{3}$        & $^1$S$^{\rm o}_{0}$    & 133 348  & 0.0000   \\
19  & $6s^{2}7s6p$      & $^1$P$^{\rm o}_{1}$    & 134 305  & 1.1032   \\
20  & $6s^{2}6p7p$      & $^3$P$_{1}$    & 135 419  & 1.4520   \\
\hline \hline
    \end{tabular}
\end{table}

\begin{table}[htpb]
    \caption{Po IV}
    \label{Po IV}
    \centering
    \begin{tabular}{ccccc}
    \hline \hline
No. & Conf.             & Term           & Energy ($\text{cm}^{-1}$) & $g_J$    \\ \hline
1   & $6s^{2}6p$        & $^2$P$^{\rm o}_{1/2}$  & 0        & 0.6667   \\
2   & $6s^{2}6p$        & $^2$P$^{\rm o}_{3/2}$  & 29 281   & 1.3333   \\
3   & $6s6p^{2}$        & $^4$P$_{1/2}$  & 83 185   & 2.4794   \\
4   & $6s6p^{2}$        & $^4$P$_{3/2}$  & 102 281  & 1.6948   \\
5   & $6s6p^{2}$        & $^4$D$_{5/2}$  & 109 905  & 1.4308   \\
6   & $6s6p^{2}$        & $^2$D$_{3/2}$  & 119 854  & 0.8814   \\
7   & $6s6p^{2}$        & $^2$P$_{1/2}$  & 128 284  & 1.0728   \\
8   & $6s^{2}6d$        & $^4$D$_{5/2}$  & 132 263  & 1.3106   \\
9   & $6s^{2}7s$        & $^2$S$_{1/2}$  & 133 449  & 2.0000   \\
10  & $6s^{2}6d$        & $^2$D$_{3/2}$  & 137 865  & 0.8625   \\
11  & $6s6p^{2}$        & $^2$D$_{5/2}$  & 146 807  & 1.2586   \\
12  & $6s6p^{2}$        & $^4$D$_{3/2}$  & 160 857  & 1.2279   \\
13  & $6s6p^{2}$        & $^2$S$_{1/2}$  & 160 965  & 1.7815   \\
14  & $6s^{2}7p$        & $^2$P$^{\rm o}_{1/2}$  & 161 894  & 0.6669   \\
15  & $6s^{2}7p$        & $^2$P$^{\rm o}_{3/2}$  & 169 992  & 1.3333   \\
16  & $6s^{2}5f$        & $^2$F$^{\rm o}_{7/2}$  & 170 209  & 1.1430   \\
17  & $6s^{2}5f$        & $^2$F$^{\rm o}_{5/2}$  & 170 317  & 0.8573   \\
18  & $6s6p6d$          & $^4$D$^{\rm o}_{3/2}$  & 206 356  & 1.0155   \\
19  & $6s^{2}7d$        & $^2$D$_{5/2}$  & 208 706  & 1.2000   \\
20  & $6s6p6d$          & $^2$D$^{\rm o}_{3/2}$  & 211 253  & 0.6826   \\
\hline \hline
    \end{tabular}
\end{table}

\begin{table}[htpb]
    \caption{Po V}
    \label{Po V}
    \centering
    \begin{tabular}{ccccc}
    \hline \hline
No. & Conf.             & Term           & Energy ($\text{cm}^{-1}$) & $g_J$    \\ \hline
1   & $6s^{2}$          & $^1$S$_{0}$    & 0        & 0.0000   \\
2   & $6s6p$            & $^3$P$^{\rm o}_{0}$    & 82 010   & 0.0000   \\
3   & $6s6p$            & $^3$P$^{\rm o}_{1}$    & 87 752   & 1.4377   \\
4   & $6s6p$            & $^3$P$^{\rm o}_{2}$    & 116 157  & 1.5000   \\
5   & $6s6p$            & $^1$P$^{\rm o}_{1}$    & 134 901  & 1.0623   \\
6   & $6p^{2}$          & $^1$S$_{0}$    & 189 607  & 0.0000   \\
7   & $6p^{2}$          & $^3$P$_{1}$    & 216 534  & 1.5004   \\
8   & $6p^{2}$          & $^3$D$_{2}$    & 217 185  & 1.1463   \\
9   & $6s6d$            & $^3$D$_{1}$    & 238 026  & 0.5002   \\
10  & $6s6d$            & $^3$D$_{2}$    & 238 767  & 1.1712   \\
11  & $6s6d$            & $^3$D$_{3}$    & 241 803  & 1.3333   \\
12  & $6s6d$            & $^3$D$_{2}$    & 243 018  & 1.2260   \\
13  & $6s7s$            & $^3$S$_{1}$    & 247 789  & 1.9994   \\
14  & $6s7s$            & $^1$S$_{0}$    & 252 112  & 0.0000   \\
15  & $6p^{2}$          & $^3$D$_{2}$    & 260 575  & 1.1232   \\
16  & $6s5f$            & $^3$F$^{\rm o}_{2}$    & 264 180  & 0.6667   \\
17  & $6s5f$            & $^3$F$^{\rm o}_{3}$    & 264 292  & 1.0833   \\
18  & $6p^{2}$          & $^1$S$_{0}$    & 269 075  & 0.0000   \\
19  & $6s5f$            & $^1$F$^{\rm o}_{3}$    & 270 173  & 1.0001   \\
20  & $6s7p$            & $^1$S$^{\rm o}_{0}$    & 281 195  & 0.0000   \\
\hline \hline
    \end{tabular}
\end{table}

\begin{table}[htpb]
    \caption{Po VI}
    \label{Po VI}
    \centering
    \begin{tabular}{ccccc}
    \hline \hline
No. & Conf.             & Term           & Energy ($\text{cm}^{-1}$) & $g_J$    \\ \hline
1   & $6s$              & $^2$S$_{1/2}$  & 0        & 2.0000   \\
2   & $6p$              & $^2$P$^{\rm o}_{1/2}$  & 99 361   & 0.6667   \\
3   & $6p$              & $^2$P$^{\rm o}_{3/2}$  & 136 739  & 1.3333   \\
4   & $6d$              & $^2$D$_{3/2}$  & 264 081  & 0.8000   \\
5   & $6d$              & $^2$D$_{5/2}$  & 268 645  & 1.2000   \\
6   & $5f$              & $^2$F$^{\rm o}_{5/2}$  & 277 071  & 0.8571   \\
7   & $5f$              & $^2$F$^{\rm o}_{7/2}$  & 277 542  & 1.1429   \\
8   & $7s$              & $^2$S$_{1/2}$  & 285 750  & 2.0000   \\
9   & $7p$              & $^2$P$^{\rm o}_{1/2}$  & 322 230  & 0.6667   \\
10  & $7p$              & $^2$P$^{\rm o}_{3/2}$  & 335 492  & 1.3333   \\
11  & $7d$              & $^2$D$_{3/2}$  & 388 974  & 0.8000   \\
12  & $6f$              & $^2$F$^{\rm o}_{5/2}$  & 390 833  & 0.8571   \\
13  & $7d$              & $^2$D$_{5/2}$  & 391 033  & 1.2000   \\
14  & $6f$              & $^2$F$^{\rm o}_{7/2}$  & 391 298  & 1.1429   \\
15  & $8s$              & $^2$S$_{1/2}$  & 396 846  & 2.0000   \\
16  & $5g$              & $^2$G$_{7/2}$  & 412 476  & 0.8889   \\
17  & $8p$              & $^2$P$^{\rm o}_{1/2}$  & 414 875  & 0.6667   \\
18  & $8p$              & $^2$P$^{\rm o}_{3/2}$  & 421 564  & 1.3333   \\
19  & $8d$              & $^2$D$_{3/2}$  & 449 717  & 0.8000   \\
20  & $7f$              & $^2$F$^{\rm o}_{5/2}$  & 449 869  & 0.8571   \\
\hline \hline
    \end{tabular}
\end{table}

\newpage

\begin{table}[htpb]
    \caption{At II}
    \label{At II}
    \centering
    \begin{tabular}{ccccc}
    \hline \hline
No. & Conf.             & Term           & Energy ($\text{cm}^{-1}$) & $g_J$    \\ \hline
1   & $6s^{2}6p^{4}$    & $^3$P$_{2}$    & 0        & 1.3855   \\
2   & $6s^{2}6p^{4}$    & $^1$S$_{0}$    & 9 890    & 0.0000   \\
3   & $6s^{2}6p^{4}$    & $^3$P$_{1}$    & 25 686   & 1.5005   \\
4   & $6s^{2}6p^{4}$    & $^3$D$_{2}$    & 31 347   & 1.1148   \\
5   & $6s^{2}6p^{4}$    & $^1$S$_{0}$    & 62 926   & 0.0000   \\
6   & $6s^{2}7s6p^{3}$  & $^5$P$^{\rm o}_{2}$    & 72 809   & 1.6751   \\
7   & $6s^{2}7s6p^{3}$  & $^3$P$^{\rm o}_{1}$    & 74 658   & 1.3266   \\
8   & $6s^{2}6p^{3}6d$  & $^3$D$^{\rm o}_{2}$    & 81 142   & 1.0888   \\
9   & $6s^{2}6p^{3}6d$  & $^3$D$^{\rm o}_{3}$    & 81 791   & 1.3033   \\
10  & $6s^{2}6p^{3}6d$  & $^5$F$^{\rm o}_{4}$    & 82 502   & 1.3523   \\
11  & $6s^{2}6p^{3}6d$  & $^1$S$^{\rm o}_{0}$    & 82 664   & 0.0000   \\
12  & $6s^{2}6p^{3}6d$  & $^3$P$^{\rm o}_{1}$    & 82 776   & 1.3286   \\
13  & $6s^{2}6p^{3}6d$  & $^3$D$^{\rm o}_{2}$    & 84 646   & 1.3231   \\
14  & $6s^{2}6p^{3}6d$  & $^7$G$^{\rm o}_{3}$    & 91 225   & 1.1304   \\
15  & $6s^{2}6p^{3}7p$  & $^3$P$_{1}$    & 91 698   & 1.5464   \\
16  & $6s^{2}6p^{3}6d$  & $^1$P$^{\rm o}_{1}$    & 92 195   & 0.9604   \\
17  & $6s^{2}6p^{3}7p$  & $^3$P$_{2}$    & 92 312   & 1.3719   \\
18  & $6s^{2}6p^{3}7p$  & $^5$D$_{3}$    & 94 732   & 1.4515   \\
19  & $6s^{2}6p^{3}7p$  & $^3$P$_{1}$    & 95 666   & 1.6594   \\
20  & $6s^{2}7s6p^{3}$  & $^3$P$^{\rm o}_{2}$    & 95 962   & 1.4894   \\
\hline \hline
    \end{tabular}
\end{table}

\begin{table}[htpb]
    \caption{At III}
    \label{At III}
    \centering
    \begin{tabular}{ccccc}
    \hline \hline
No. & Conf.             & Term           & Energy ($\text{cm}^{-1}$) & $g_J$    \\ \hline
1   & $6s^{2}6p^{3}$    & $^2$P$^{\rm o}_{3/2}$  & 0        & 1.5208   \\
2   & $6s^{2}6p^{3}$    & $^2$P$^{\rm o}_{3/2}$  & 24 339   & 1.3451   \\
3   & $6s^{2}6p^{3}$    & $^2$D$^{\rm o}_{5/2}$  & 30 187   & 1.2000   \\
4   & $6s^{2}6p^{3}$    & $^2$P$^{\rm o}_{1/2}$  & 38 976   & 0.6666   \\
5   & $6s^{2}6p^{3}$    & $^2$P$^{\rm o}_{3/2}$  & 64 321   & 1.2676   \\
6   & $6s^{2}7s6p^{2}$  & $^2$S$_{1/2}$  & 93 135   & 2.0765   \\
7   & $6s^{2}6p^{2}6d$  & $^2$D$_{3/2}$  & 96 145   & 0.7895   \\
8   & $6s^{2}6p^{2}6d$  & $^6$F$_{5/2}$  & 99 904   & 1.2956   \\
9   & $6s6p^{4}$        & $^4$D$_{5/2}$  & 110 684  & 1.3837   \\
10  & $6s^{2}6p^{2}6d$  & $^6$F$_{3/2}$  & 116 460  & 1.0601   \\
11  & $6s^{2}6p^{2}7p$  & $^2$P$^{\rm o}_{1/2}$  & 116 610  & 0.6737   \\
12  & $6s^{2}7s6p^{2}$  & $^4$P$_{3/2}$  & 118 634  & 1.6789   \\
13  & $6s^{2}7s6p^{2}$  & $^2$P$_{1/2}$  & 119 062  & 1.1839   \\
14  & $6s^{2}6p^{2}6d$  & $^4$F$_{5/2}$  & 120 692  & 0.9523   \\
15  & $6s6p^{4}$        & $^2$S$_{1/2}$  & 120 954  & 2.2238   \\
16  & $6s^{2}6p^{2}6d$  & $^4$F$_{7/2}$  & 121 939  & 1.2755   \\
17  & $6s^{2}6p^{2}6d$  & $^2$P$_{1/2}$  & 122 438  & 0.3984   \\
18  & $6s^{2}6p^{2}6d$  & $^6$F$_{3/2}$  & 123 078  & 1.0485   \\
19  & $6s^{2}7s6p^{2}$  & $^4$D$_{5/2}$  & 123 580  & 1.3779   \\
20  & $6s^{2}6p^{2}6d$  & $^2$F$_{7/2}$  & 124 754  & 1.1692   \\
\hline \hline
    \end{tabular}
\end{table}

\begin{table}[htpb]
    \caption{At IV}
    \label{At IV}
    \centering
    \begin{tabular}{ccccc}
    \hline \hline
No. & Conf.             & Term           & Energy ($\text{cm}^{-1}$) & $g_J$    \\ \hline
1   & $6s^{2}6p^{2}$    & $^1$S$_{0}$    & 0        & 0.0000   \\
2   & $6s^{2}6p^{2}$    & $^3$P$_{1}$    & 27 414   & 1.4999   \\
3   & $6s^{2}6p^{2}$    & $^3$D$_{2}$    & 32 524   & 1.2253   \\
4   & $6s^{2}6p^{2}$    & $^3$D$_{2}$    & 64 774   & 1.2747   \\
5   & $6s^{2}6p^{2}$    & $^1$S$_{0}$    & 77 897   & 0.0000   \\
6   & $6s6p^{3}$        & $^5$P$^{\rm o}_{2}$    & 115 945  & 1.7463   \\
7   & $6s6p^{3}$        & $^1$P$^{\rm o}_{1}$    & 130 189  & 0.8404   \\
8   & $6s^{2}6p6d$      & $^3$F$^{\rm o}_{2}$    & 138 425  & 0.7883   \\
9   & $6s6p^{3}$        & $^3$P$^{\rm o}_{2}$    & 139 485  & 1.4110   \\
10  & $6s^{2}6p6d$      & $^5$F$^{\rm o}_{3}$    & 144 572  & 1.2034   \\
11  & $6s^{2}7s6p$      & $^1$S$^{\rm o}_{0}$    & 144 911  & 0.0000   \\
12  & $6s^{2}7s6p$      & $^3$P$^{\rm o}_{1}$    & 145 715  & 1.3434   \\
13  & $6s^{2}6p6d$      & $^1$P$^{\rm o}_{1}$    & 153 291  & 0.9252   \\
14  & $6s^{2}6p6d$      & $^3$D$^{\rm o}_{2}$    & 154 824  & 1.1893   \\
15  & $6s6p^{3}$        & $^5$F$^{\rm o}_{3}$    & 155 665  & 1.2325   \\
16  & $6s6p^{3}$        & $^1$S$^{\rm o}_{0}$    & 162 027  & 0.0000   \\
17  & $6s6p^{3}$        & $^3$P$^{\rm o}_{1}$    & 165 260  & 1.2610   \\
18  & $6s^{2}6p7p$      & $^3$D$_{1}$    & 173 686  & 0.6718   \\
19  & $6s^{2}6p5f$      & $^5$G$_{3}$    & 177 765  & 0.8359   \\
20  & $6s^{2}6p7p$      & $^1$S$_{0}$    & 178 535  & 0.0000   \\
\hline \hline
    \end{tabular}
\end{table}

\begin{table}[htpb]
    \caption{At V}
    \label{At V}
    \centering
    \begin{tabular}{ccccc}
    \hline \hline
No. & Conf.             & Term           & Energy ($\text{cm}^{-1}$) & $g_J$    \\ \hline
1   & $6s^{2}6p$        & $^2$P$^{\rm o}_{1/2}$  & 0        & 0.6667   \\
2   & $6s^{2}6p$        & $^2$P$^{\rm o}_{3/2}$  & 37 275   & 1.3333   \\
3   & $6s6p^{2}$        & $^4$P$_{1/2}$  & 94 874   & 2.4460   \\
4   & $6s6p^{2}$        & $^4$P$_{3/2}$  & 120 201  & 1.6923   \\
5   & $6s6p^{2}$        & $^4$D$_{5/2}$  & 128 971  & 1.4233   \\
6   & $6s6p^{2}$        & $^2$D$_{3/2}$  & 140 916  & 0.9019   \\
7   & $6s6p^{2}$        & $^2$P$_{1/2}$  & 148 155  & 1.0899   \\
8   & $6s6p^{2}$        & $^4$D$_{5/2}$  & 160 043  & 1.3335   \\
9   & $6s^{2}6d$        & $^2$D$_{3/2}$  & 166 521  & 0.8622   \\
10  & $6s^{2}7s$        & $^2$S$_{1/2}$  & 173 726  & 1.9947   \\
11  & $6s^{2}6d$        & $^2$D$_{5/2}$  & 176 417  & 1.2431   \\
12  & $6s^{2}5f$        & $^2$F$^{\rm o}_{5/2}$  & 189 117  & 0.8572   \\
13  & $6s^{2}5f$        & $^2$F$^{\rm o}_{7/2}$  & 189 403  & 1.1429   \\
14  & $6s6p^{2}$        & $^4$D$_{3/2}$  & 189 417  & 1.2101   \\
15  & $6s6p^{2}$        & $^2$S$_{1/2}$  & 189 616  & 1.8028   \\
16  & $6s^{2}7p$        & $^2$P$^{\rm o}_{1/2}$  & 209 008  & 0.6670   \\
17  & $6s^{2}7p$        & $^2$P$^{\rm o}_{3/2}$  & 220 236  & 1.3330   \\
18  & $6p^{3}$          & $^4$D$^{\rm o}_{3/2}$  & 239 964  & 1.2187   \\
19  & $6s6p6d$          & $^4$F$^{\rm o}_{3/2}$  & 249 401  & 0.5658   \\
20  & $6s6p6d$          & $^4$F$^{\rm o}_{5/2}$  & 253 313  & 1.0720   \\
\hline \hline
    \end{tabular}
\end{table}

\begin{table}[htpb]
    \caption{At VI}
    \label{At VI}
    \centering
    \begin{tabular}{ccccc}
    \hline \hline
No. & Conf.             & Term           & Energy ($\text{cm}^{-1}$) & $g_J$    \\ \hline
1   & $6s^{2}$          & $^1$S$_{0}$    & 0        & 0.0000   \\
2   & $6s6p$            & $^3$P$^{\rm o}_{0}$    & 92 353   & 0.0000   \\
3   & $6s6p$            & $^3$P$^{\rm o}_{1}$    & 98 860   & 1.4303   \\
4   & $6s6p$            & $^3$P$^{\rm o}_{2}$    & 134 698  & 1.5000   \\
5   & $6s6p$            & $^1$P$^{\rm o}_{1}$    & 154 856  & 1.0697   \\
6   & $6p^{2}$          & $^1$S$_{0}$    & 212 464  & 0.0000   \\
7   & $6p^{2}$          & $^3$P$_{1}$    & 246 771  & 1.5002   \\
8   & $6p^{2}$          & $^3$D$_{2}$    & 248 178  & 1.1583   \\
9   & $6s6d$            & $^3$D$_{1}$    & 277 272  & 0.5001   \\
10  & $6s6d$            & $^3$D$_{2}$    & 277 957  & 1.1747   \\
11  & $6s6d$            & $^3$D$_{2}$    & 282 355  & 1.2241   \\
12  & $6s6d$            & $^3$D$_{3}$    & 282 443  & 1.3333   \\
13  & $6s5f$            & $^3$F$^{\rm o}_{2}$    & 285 273  & 0.6667   \\
14  & $6s5f$            & $^3$F$^{\rm o}_{3}$    & 285 596  & 1.0832   \\
15  & $6s5f$            & $^3$F$^{\rm o}_{4}$    & 285 925  & 1.2500   \\
16  & $6s5f$            & $^1$F$^{\rm o}_{3}$    & 293 762  & 1.0002   \\
17  & $6s7s$            & $^3$S$_{1}$    & 299 794  & 1.9997   \\
18  & $6p^{2}$          & $^3$D$_{2}$    & 301 257  & 1.1093   \\
19  & $6s7s$            & $^1$S$_{0}$    & 302 114  & 0.0000   \\
20  & $6p^{2}$          & $^1$S$_{0}$    & 311 378  & 0.0000   \\
\hline \hline
    \end{tabular}
\end{table}

\begin{table}[htpb]
    \caption{At VII}
    \label{At VII}
    \centering
    \begin{tabular}{ccccc}
    \hline \hline
No. & Conf.             & Term           & Energy ($\text{cm}^{-1}$) & $g_J$    \\ \hline
1   & $6s$              & $^2$S$_{1/2}$  & 0        & 2.0000   \\
2   & $6p$              & $^2$P$^{\rm o}_{1/2}$  & 110 532  & 0.6667   \\
3   & $6p$              & $^2$P$^{\rm o}_{3/2}$  & 156 249  & 1.3333   \\
4   & $5f$              & $^2$F$^{\rm o}_{5/2}$  & 293 177  & 0.8571   \\
5   & $5f$              & $^2$F$^{\rm o}_{7/2}$  & 294 304  & 1.1429   \\
6   & $6d$              & $^2$D$_{3/2}$  & 302 953  & 0.8000   \\
7   & $6d$              & $^2$D$_{5/2}$  & 308 934  & 1.2000   \\
8   & $7s$              & $^2$S$_{1/2}$  & 338 854  & 2.0000   \\
9   & $7p$              & $^2$P$^{\rm o}_{1/2}$  & 381 358  & 0.6667   \\
10  & $7p$              & $^2$P$^{\rm o}_{3/2}$  & 398 290  & 1.3333   \\
11  & $6f$              & $^2$F$^{\rm o}_{5/2}$  & 454 374  & 0.8571   \\
12  & $6f$              & $^2$F$^{\rm o}_{7/2}$  & 455 177  & 1.1429   \\
13  & $7d$              & $^2$D$_{3/2}$  & 461 790  & 0.8000   \\
14  & $7d$              & $^2$D$_{5/2}$  & 464 529  & 1.2000   \\
15  & $8s$              & $^2$S$_{1/2}$  & 476 199  & 2.0000   \\
16  & $5g$              & $^2$G$_{9/2}$  & 482 854  & 1.1111   \\
17  & $5g$              & $^2$G$_{7/2}$  & 482 921  & 0.8889   \\
18  & $8p$              & $^2$P$^{\rm o}_{1/2}$  & 497 502  & 0.6667   \\
19  & $8p$              & $^2$P$^{\rm o}_{3/2}$  & 506 093  & 1.3333   \\
20  & $7f$              & $^2$F$^{\rm o}_{5/2}$  & 535 324  & 0.8571   \\
\hline \hline
    \end{tabular}
\end{table}

\begin{table}[htpb]
    \caption{Lv II}
    \label{Lv II}
    \centering
    \begin{tabular}{ccccc}
    \hline \hline
No. & Conf.             & Term           & Energy ($\text{cm}^{-1}$) & $g_J$    \\ \hline
1   & $7s^{2}7p^{3}$    & $^2$P$^{\rm o}_{3/2}$  & 0        & 1.3868   \\
2   & $7s^{2}8s7p^{2}$  & $^2$S$_{1/2}$  & 39 298   & 2.0324   \\
3   & $7s^{2}7p^{3}$    & $^2$P$^{\rm o}_{3/2}$  & 53 559   & 1.4501   \\
4   & $7s^{2}7p^{2}8p$  & $^2$P$^{\rm o}_{1/2}$  & 55 706   & 0.6673   \\
5   & $7s^{2}7p^{2}7d$  & $^2$D$_{3/2}$  & 57 417   & 0.7924   \\
6   & $7s^{2}7p^{3}$    & $^2$D$^{\rm o}_{5/2}$  & 58 490   & 1.1994   \\
7   & $7s^{2}7p^{2}7d$  & $^2$D$_{5/2}$  & 60 204   & 1.2023   \\
8   & $7s^{2}7p^{3}$    & $^2$P$^{\rm o}_{1/2}$  & 65 119   & 0.6663   \\
9   & $7s^{2}7p^{2}9p$  & $^2$P$^{\rm o}_{3/2}$  & 66 217   & 1.3360   \\
10  & $7s^{2}10s7p^{2}$ & $^2$S$_{1/2}$  & 75 484   & 2.0053   \\
11  & $7s^{2}7p^{2}7f$  & $^2$F$^{\rm o}_{5/2}$  & 81 477   & 0.8560   \\
12  & $7s^{2}7p^{2}7f$  & $^2$F$^{\rm o}_{7/2}$  & 81 585   & 1.1434   \\
13  & $7s^{2}7p^{2}10p$ & $^2$P$^{\rm o}_{1/2}$  & 81 818   & 0.6669   \\
14  & $7s^{2}7p^{2}9d$  & $^2$D$_{3/2}$  & 82 249   & 0.7989   \\
15  & $7s^{2}7p^{2}9d$  & $^2$D$_{5/2}$  & 82 976   & 1.2005   \\
16  & $7s^{2}7p^{2}10p$ & $^2$P$^{\rm o}_{3/2}$  & 85 078   & 1.3341   \\
17  & $7s^{2}12s7p^{2}$ & $^2$S$_{1/2}$  & 86 428   & 2.0023   \\
18  & $7s^{2}7p^{2}7g$  & $^2$G$_{7/2}$  & 88 472   & 0.8889   \\
19  & $7s^{2}7p^{2}7g$  & $^2$G$_{9/2}$  & 88 760   & 1.1111   \\
20  & $7s^{2}7p^{2}10f$ & $^2$F$^{\rm o}_{5/2}$  & 90 475   & 0.8561   \\
\hline \hline
    \end{tabular}
\end{table}

\begin{table}[htpb]
    \caption{Lv III}
    \label{Lv III}
    \centering
    \begin{tabular}{ccccc}
    \hline \hline
No. & Conf.             & Term           & Energy ($\text{cm}^{-1}$) & $g_J$    \\ \hline
1   & $7s^{2}7p^{2}$    & $^1$S$_{0}$    & 0        & 0.0000   \\
2   & $7s^{2}7p^{2}$    & $^3$P$_{1}$    & 60 881   & 1.4996   \\
3   & $7s^{2}7p^{2}$    & $^3$D$_{2}$    & 65 127   & 1.1882   \\
4   & $7s^{2}8s7p$      & $^1$S$^{\rm o}_{0}$    & 120 149  & 0.0000   \\
5   & $7s^{2}8s7p$      & $^3$P$^{\rm o}_{1}$    & 120 569  & 1.3375   \\
6   & $7s^{2}7p^{2}$    & $^3$D$_{2}$    & 132 446  & 1.3106   \\
7   & $7s^{2}7p7d$      & $^3$F$^{\rm o}_{2}$    & 136 141  & 0.7821   \\
8   & $7s^{2}7p7d$      & $^1$P$^{\rm o}_{1}$    & 139 505  & 0.8222   \\
9   & $7s^{2}7p7d$      & $^3$D$^{\rm o}_{2}$    & 140 057  & 1.2871   \\
10  & $7s^{2}7p8p$      & $^3$D$_{1}$    & 141 536  & 0.6677   \\
11  & $7s^{2}7p7d$      & $^3$F$^{\rm o}_{3}$    & 141 988  & 1.1170   \\
12  & $7s^{2}7p^{2}$    & $^1$S$_{0}$    & 142 194  & 0.0000   \\
13  & $7s^{2}7p8p$      & $^1$S$_{0}$    & 145 555  & 0.0000   \\
14  & $7s7p^{3}$        & $^3$P$^{\rm o}_{2}$    & 154 116  & 1.5708   \\
15  & $7s^{2}7p8p$      & $^3$P$_{1}$    & 157 196  & 1.4878   \\
16  & $7s^{2}7p8p$      & $^3$D$_{2}$    & 157 750  & 1.1701   \\
17  & $7s7p^{3}$        & $^1$P$^{\rm o}_{1}$    & 163 461  & 1.0858   \\
18  & $7s^{2}7p6f$      & $^3$G$_{3}$    & 163 778  & 0.8270   \\
19  & $7s^{2}7p6f$      & $^5$F$_{3}$    & 166 831  & 1.1924   \\
20  & $7s^{2}7p6f$      & $^1$D$_{2}$    & 167 525  & 0.8757   \\
\hline \hline
    \end{tabular}
\end{table}

\begin{table}[htpb]
    \caption{Lv IV}
    \label{Lv IV}
    \centering
    \begin{tabular}{ccccc}
    \hline \hline
No. & Conf.             & Term           & Energy ($\text{cm}^{-1}$) & $g_J$    \\ \hline
1   & $7s^{2}7p$        & $^2$P$^{\rm o}_{1/2}$  & 0        & 0.6667   \\
2   & $7s^{2}7p$        & $^2$P$^{\rm o}_{3/2}$  & 76 304   & 1.3334   \\
3   & $7s7p^{2}$        & $^2$S$_{1/2}$  & 95 232   & 2.1994   \\
4   & $7s^{2}8s$        & $^2$S$_{1/2}$  & 150 819  & 1.9906   \\
5   & $7s7p^{2}$        & $^4$P$_{3/2}$  & 159 714  & 1.5473   \\
6   & $7s^{2}7d$        & $^2$D$_{3/2}$  & 165 027  & 0.9333   \\
7   & $7s7p^{2}$        & $^4$D$_{5/2}$  & 165 069  & 1.2890   \\
8   & $7s^{2}7d$        & $^2$D$_{5/2}$  & 174 863  & 1.2835   \\
9   & $7s7p^{2}$        & $^2$P$_{1/2}$  & 176 761  & 1.2231   \\
10  & $7s7p^{2}$        & $^2$D$_{3/2}$  & 178 298  & 0.9807   \\
11  & $7s^{2}8p$        & $^2$P$^{\rm o}_{1/2}$  & 178 837  & 0.6670   \\
12  & $7s^{2}6f$        & $^2$F$^{\rm o}_{5/2}$  & 185 773  & 0.8576   \\
13  & $7s^{2}6f$        & $^2$F$^{\rm o}_{7/2}$  & 188 417  & 1.1429   \\
14  & $7s^{2}8p$        & $^2$P$^{\rm o}_{3/2}$  & 198 642  & 1.3334   \\
15  & $7s^{2}9s$        & $^2$S$_{1/2}$  & 224 682  & 2.0006   \\
16  & $7s^{2}8d$        & $^2$D$_{3/2}$  & 232 165  & 0.8029   \\
17  & $7s^{2}8d$        & $^2$D$_{5/2}$  & 234 141  & 1.2101   \\
18  & $7s^{2}9p$        & $^2$P$^{\rm o}_{1/2}$  & 236 959  & 0.7341   \\
19  & $7s^{2}7f$        & $^2$F$^{\rm o}_{5/2}$  & 238 099  & 0.8639   \\
20  & $7s^{2}7f$        & $^2$F$^{\rm o}_{7/2}$  & 240 011  & 1.1429   \\
\hline \hline
    \end{tabular}
\end{table}

\begin{table}[htpb]
    \caption{Lv V}
    \label{Lv V}
    \centering
    \begin{tabular}{ccccc}
    \hline \hline
No. & Conf.             & Term           & Energy ($\text{cm}^{-1}$) & $g_J$    \\ \hline
1   & $7s^{2}$          & $^1$S$_{0}$    & 0        & 0.0000   \\
2   & $7s7p$            & $^3$P$^{\rm o}_{0}$    & 90 089   & 0.0000   \\
3   & $7s7p$            & $^3$P$^{\rm o}_{1}$    & 96 927   & 1.3755   \\
4   & $7s7p$            & $^3$P$^{\rm o}_{2}$    & 178 355  & 1.5000   \\
5   & $7s7p$            & $^1$P$^{\rm o}_{1}$    & 189 943  & 1.1244   \\
6   & $7p^{2}$          & $^1$S$_{0}$    & 202 863  & 0.0000   \\
7   & $7s8s$            & $^3$S$_{1}$    & 274 050  & 1.9680   \\
8   & $7s8s$            & $^1$S$_{0}$    & 279 070  & 0.0000   \\
9   & $7p^{2}$          & $^3$D$_{2}$    & 280 718  & 1.0898   \\
10  & $7p^{2}$          & $^3$P$_{1}$    & 282 832  & 1.3800   \\
11  & $7s7d$            & $^3$D$_{1}$    & 284 724  & 0.6520   \\
12  & $7s7d$            & $^3$D$_{2}$    & 286 047  & 1.1683   \\
13  & $7s6f$            & $^3$F$^{\rm o}_{2}$    & 290 152  & 0.6670   \\
14  & $7s7d$            & $^3$D$_{3}$    & 291 502  & 1.3333   \\
15  & $7s6f$            & $^3$F$^{\rm o}_{3}$    & 291 773  & 1.0804   \\
16  & $7s6f$            & $^3$F$^{\rm o}_{4}$    & 293 353  & 1.2500   \\
17  & $7s7d$            & $^3$D$_{2}$    & 297 222  & 1.0966   \\
18  & $7s6f$            & $^1$F$^{\rm o}_{3}$    & 299 393  & 1.0030   \\
19  & $7s8p$            & $^1$S$^{\rm o}_{0}$    & 307 649  & 0.0000   \\
20  & $7s8p$            & $^3$P$^{\rm o}_{1}$    & 308 154  & 1.3455   \\
\hline \hline
    \end{tabular}
\end{table}

\begin{table}[t]
    \caption{Lv VI}
    \label{Lv VI}
    \centering
    \begin{tabular}{ccccc}
    \hline \hline
No. & Conf.             & Term           & Energy ($\text{cm}^{-1}$) & $g_J$    \\ \hline
1   & $7s$              & $^2$S$_{1/2}$  & 0        & 2.0000   \\
2   & $7p$              & $^2$P$^{\rm o}_{1/2}$  & 104 747  & 0.6667   \\
3   & $7p$              & $^2$P$^{\rm o}_{3/2}$  & 200 696  & 1.3333   \\
4   & $6f$              & $^2$F$^{\rm o}_{5/2}$  & 307 694  & 0.8571   \\
5   & $6f$              & $^2$F$^{\rm o}_{7/2}$  & 309 985  & 1.1429   \\
6   & $8s$              & $^2$S$_{1/2}$  & 312 384  & 2.0000   \\
7   & $7d$              & $^2$D$_{3/2}$  & 314 134  & 0.8000   \\
8   & $7d$              & $^2$D$_{5/2}$  & 322 793  & 1.2000   \\
9   & $8p$              & $^2$P$^{\rm o}_{1/2}$  & 348 592  & 0.6667   \\
10  & $8p$              & $^2$P$^{\rm o}_{3/2}$  & 380 698  & 1.3333   \\
11  & $9s$              & $^2$S$_{1/2}$  & 426 042  & 2.0000   \\
12  & $7f$              & $^2$F$^{\rm o}_{5/2}$  & 426 119  & 0.8571   \\
13  & $7f$              & $^2$F$^{\rm o}_{7/2}$  & 427 401  & 1.1429   \\
14  & $8d$              & $^2$D$_{3/2}$  & 429 004  & 0.8000   \\
15  & $8d$              & $^2$D$_{5/2}$  & 432 967  & 1.2000   \\
16  & $5g$              & $^2$G$_{7/2}$  & 437 108  & 0.8889   \\
17  & $9p$              & $^2$P$^{\rm o}_{1/2}$  & 443 795  & 0.6667   \\
18  & $9p$              & $^2$P$^{\rm o}_{3/2}$  & 459 827  & 1.3333   \\
19  & $10s$             & $^2$S$_{1/2}$  & 483 054  & 2.0000   \\
20  & $8f$              & $^2$F$^{\rm o}_{5/2}$  & 483 763  & 0.8571   \\
\hline \hline
    \end{tabular}
\end{table}

\begin{table}[htpb]
    \caption{Ts II}
    \label{Ts II}
    \centering
    \begin{tabular}{ccccc}
    \hline \hline
No. & Conf.             & Term           & Energy ($\text{cm}^{-1}$) & $g_J$    \\ \hline
1   & $7s^{2}7p^{4}$    & $^3$P$_{2}$    & 0        & 1.3538   \\
2   & $7s^{2}7p^{4}$    & $^1$S$_{0}$    & 8 415    & 0.0000   \\
3   & $7s^{2}8s7p^{3}$  & $^3$P$^{\rm o}_{2}$    & 47 458   & 1.5445   \\
4   & $7s^{2}8s7p^{3}$  & $^1$P$^{\rm o}_{1}$    & 50 201   & 1.1960   \\
5   & $7s^{2}7p^{3}7d$  & $^1$D$^{\rm o}_{2}$    & 64 901   & 0.9969   \\
6   & $7s^{2}7p^{3}7d$  & $^7$G$^{\rm o}_{3}$    & 66 610   & 1.1973   \\
7   & $7s^{2}7p^{3}7d$  & $^3$F$^{\rm o}_{4}$    & 67 228   & 1.2722   \\
8   & $7s^{2}7p^{3}7d$  & $^1$S$^{\rm o}_{0}$    & 67 509   & 0.0000   \\
9   & $7s^{2}7p^{3}7d$  & $^1$P$^{\rm o}_{1}$    & 67 748   & 1.2044   \\
10  & $7s^{2}7p^{3}8p$  & $^3$P$_{1}$    & 67 831   & 1.5241   \\
11  & $7s^{2}7p^{3}8p$  & $^3$D$_{2}$    & 68 378   & 1.2150   \\
12  & $7s^{2}7p^{4}$    & $^3$P$_{1}$    & 68 806   & 1.4354   \\
13  & $7s^{2}7p^{3}7d$  & $^3$D$^{\rm o}_{2}$    & 69 756   & 1.3167   \\
14  & $7s^{2}7p^{4}$    & $^3$D$_{2}$    & 72 547   & 1.1504   \\
15  & $7s^{2}7p^{3}8d$  & $^7$G$^{\rm o}_{3}$    & 73 768   & 1.1424   \\
16  & $7s^{2}7p^{3}8d$  & $^1$P$^{\rm o}_{1}$    & 75 745   & 0.9883   \\
17  & $7s^{2}7p^{3}9p$  & $^3$D$_{3}$    & 76 493   & 1.3613   \\
18  & $7s^{2}7p^{3}9p$  & $^3$P$_{1}$    & 76 921   & 1.4675   \\
19  & $7s^{2}7p^{3}9p$  & $^3$P$_{2}$    & 79 396   & 1.3391   \\
20  & $7s^{2}7p^{3}9p$  & $^1$S$_{0}$    & 81 599   & 0.0000   \\
\hline \hline
    \end{tabular}
\end{table}

\begin{table}[htpb]
    \caption{Ts III}
    \label{Ts III}
    \centering
    \begin{tabular}{ccccc}
    \hline \hline
No. & Conf.             & Term           & Energy ($\text{cm}^{-1}$) & $g_J$    \\ \hline
1   & $7s^{2}7p^{3}$    & $^2$P$^{\rm o}_{3/2}$  & 0        & 1.3796   \\
2   & $7s^{2}8s7p^{2}$  & $^2$S$_{1/2}$  & 61 438   & 2.0348   \\
3   & $7s^{2}7p^{3}$    & $^2$P$^{\rm o}_{3/2}$  & 71 046   & 1.4519   \\
4   & $7s^{2}7p^{3}$    & $^2$D$^{\rm o}_{5/2}$  & 76 097   & 1.1976   \\
5   & $7s^{2}7p^{2}7d$  & $^2$D$_{3/2}$  & 78 615   & 0.7909   \\
6   & $7s^{2}7p^{2}8p$  & $^2$P$^{\rm o}_{1/2}$  & 83 057   & 0.6654   \\
7   & $7s^{2}7p^{2}7d$  & $^2$D$_{5/2}$  & 84 137   & 1.2034   \\
8   & $7s^{2}7p^{2}8p$  & $^2$P$^{\rm o}_{1/2}$  & 85 099   & 0.6671   \\
9   & $7s^{2}7p^{2}8p$  & $^2$P$^{\rm o}_{3/2}$  & 101 442  & 1.3360   \\
10  & $7s^{2}7p^{2}6f$  & $^2$F$^{\rm o}_{5/2}$  & 108 508  & 0.8499   \\
11  & $7s^{2}7p^{2}6f$  & $^2$F$^{\rm o}_{7/2}$  & 111 899  & 1.1451   \\
12  & $7s^{2}10s7p^{2}$ & $^2$S$_{1/2}$  & 119 814  & 2.0048   \\
13  & $7s^{2}7p^{2}9d$  & $^2$D$_{3/2}$  & 127 079  & 0.7992   \\
14  & $7s^{2}7p^{2}9d$  & $^2$D$_{5/2}$  & 128 733  & 1.2007   \\
15  & $7s^{2}7p^{2}10p$ & $^2$P$^{\rm o}_{1/2}$  & 129 570  & 0.6670   \\
16  & $7s^{2}8s7p^{2}$  & $^4$P$_{3/2}$  & 135 594  & 1.6789   \\
17  & $7s^{2}7p^{2}10p$ & $^2$P$^{\rm o}_{3/2}$  & 136 200  & 1.3339   \\
18  & $7s^{2}7p^{2}8f$  & $^2$F$^{\rm o}_{5/2}$  & 136 649  & 0.8544   \\
19  & $7s^{2}8s7p^{2}$  & $^2$P$_{1/2}$  & 138 276  & 1.3054   \\
20  & $7s^{2}7p^{2}6g$  & $^2$G$_{7/2}$  & 138 755  & 0.8894   \\
\hline \hline
    \end{tabular}
\end{table}

\begin{table}[htpb]
    \caption{Ts IV}
    \label{Ts IV}
    \centering
    \begin{tabular}{ccccc}
    \hline \hline
No. & Conf.             & Term           & Energy ($\text{cm}^{-1}$) & $g_J$    \\ \hline
1   & $7s^{2}7p^{2}$    & $^1$S$_{0}$    & 0        & 0.0000   \\
2   & $7s^{2}7p^{2}$    & $^3$P$_{1}$    & 79 085   & 1.4997   \\
3   & $7s^{2}7p^{2}$    & $^3$D$_{2}$    & 83 468   & 1.1847   \\
4   & $7s^{2}8s7p$      & $^1$S$^{\rm o}_{0}$    & 161 013  & 0.0000   \\
5   & $7s^{2}8s7p$      & $^3$P$^{\rm o}_{1}$    & 161 449  & 1.3338   \\
6   & $7s^{2}7p^{2}$    & $^3$D$_{2}$    & 169 367  & 1.3112   \\
7   & $7s^{2}7p7d$      & $^1$D$^{\rm o}_{2}$    & 173 731  & 0.8519   \\
8   & $7s^{2}7p7d$      & $^3$D$^{\rm o}_{2}$    & 176 711  & 1.3248   \\
9   & $7s^{2}7p7d$      & $^1$P$^{\rm o}_{1}$    & 177 862  & 0.8363   \\
10  & $7s^{2}7p^{2}$    & $^1$S$_{0}$    & 180 548  & 0.0000   \\
11  & $7s^{2}7p7d$      & $^3$F$^{\rm o}_{3}$    & 182 797  & 1.1174   \\
12  & $7s7p^{3}$        & $^3$P$^{\rm o}_{2}$    & 188 030  & 1.4538   \\
13  & $7s^{2}7p8p$      & $^3$D$_{1}$    & 188 781  & 0.6677   \\
14  & $7s^{2}7p6f$      & $^3$G$_{3}$    & 192 044  & 0.8260   \\
15  & $7s^{2}7p8p$      & $^1$S$_{0}$    & 193 633  & 0.0000   \\
16  & $7s7p^{3}$        & $^1$P$^{\rm o}_{1}$    & 195 275  & 1.0912   \\
17  & $7s^{2}7p6f$      & $^1$D$_{2}$    & 199 210  & 0.8795   \\
18  & $7s^{2}7p6f$      & $^5$F$_{3}$    & 200 415  & 1.1919   \\
19  & $7s^{2}7p6f$      & $^3$G$_{4}$    & 202 038  & 1.0909   \\
20  & $7s^{2}7p8p$      & $^3$P$_{1}$    & 211 694  & 1.4878   \\
\hline \hline
    \end{tabular}
\end{table}

\begin{table}[htpb]
    \caption{Ts V}
    \label{Ts V}
    \centering
    \begin{tabular}{ccccc}
    \hline \hline
No. & Conf.             & Term           & Energy ($\text{cm}^{-1}$) & $g_J$    \\ \hline
1   & $7s^{2}7p$        & $^2$P$^{\rm o}_{1/2}$  & 0        & 0.6667   \\
2   & $7s^{2}7p$        & $^2$P$^{\rm o}_{3/2}$  & 95 268   & 1.3334   \\
3   & $7s7p^{2}$        & $^2$S$_{1/2}$  & 106 110  & 2.1780   \\
4   & $7s7p^{2}$        & $^4$P$_{3/2}$  & 188 262  & 1.6210   \\
5   & $7s^{2}8s$        & $^2$S$_{1/2}$  & 192 179  & 1.9727   \\
6   & $7s7p^{2}$        & $^4$D$_{5/2}$  & 195 919  & 1.3363   \\
7   & $7s^{2}7d$        & $^2$D$_{3/2}$  & 200 459  & 0.9023   \\
8   & $7s7p^{2}$        & $^2$P$_{1/2}$  & 206 836  & 1.2561   \\
9   & $7s^{2}6f$        & $^2$F$^{\rm o}_{5/2}$  & 207 894  & 0.8574   \\
10  & $7s7p^{2}$        & $^2$D$_{3/2}$  & 209 953  & 0.9387   \\
11  & $7s^{2}7d$        & $^2$D$_{5/2}$  & 212 886  & 1.2316   \\
12  & $7s^{2}6f$        & $^2$F$^{\rm o}_{7/2}$  & 215 639  & 1.1429   \\
13  & $7s^{2}8p$        & $^2$P$^{\rm o}_{1/2}$  & 226 641  & 0.6671   \\
14  & $7s^{2}8p$        & $^2$P$^{\rm o}_{3/2}$  & 254 050  & 1.3335   \\
15  & $7s^{2}9s$        & $^2$S$_{1/2}$  & 290 084  & 2.0000   \\
16  & $7s8s7p$          & $^2$S$^{\rm o}_{1/2}$  & 291 579  & 2.3040   \\
17  & $7s7p^{2}$        & $^4$D$_{5/2}$  & 291 790  & 1.4159   \\
18  & $7s^{2}8d$        & $^2$D$_{3/2}$  & 296 256  & 0.8076   \\
19  & $7s8s7p$          & $^2$P$^{\rm o}_{3/2}$  & 297 686  & 1.2950   \\
20  & $7s^{2}7f$        & $^2$F$^{\rm o}_{5/2}$  & 298 611  & 0.8575   \\
\hline \hline
    \end{tabular}
\end{table}

\begin{table}[htpb]
    \caption{Ts VI}
    \label{Ts VI}
    \centering
    \begin{tabular}{ccccc}
    \hline \hline
No. & Conf.             & Term           & Energy ($\text{cm}^{-1}$) & $g_J$    \\ \hline
1   & $7s^{2}$          & $^1$S$_{0}$    & 0        & 0.0000   \\
2   & $7s7p$            & $^3$P$^{\rm o}_{0}$    & 99 767   & 0.0000   \\
3   & $7s7p$            & $^3$P$^{\rm o}_{1}$    & 107 432  & 1.3717   \\
4   & $7s7p$            & $^3$P$^{\rm o}_{2}$    & 207 417  & 1.5000   \\
5   & $7s7p$            & $^1$P$^{\rm o}_{1}$    & 219 893  & 1.1283   \\
6   & $7p^{2}$          & $^1$S$_{0}$    & 224 256  & 0.0000   \\
7   & $7s6f$            & $^3$F$^{\rm o}_{2}$    & 318 076  & 0.6668   \\
8   & $7s6f$            & $^3$F$^{\rm o}_{3}$    & 319 942  & 1.0791   \\
9   & $7p^{2}$          & $^3$P$_{1}$    & 321 033  & 1.5785   \\
10  & $7p^{2}$          & $^3$D$_{2}$    & 322 265  & 1.1264   \\
11  & $7s6f$            & $^3$F$^{\rm o}_{4}$    & 326 511  & 1.2500   \\
12  & $7s8s$            & $^3$S$_{1}$    & 327 745  & 1.9022   \\
13  & $7s6f$            & $^1$F$^{\rm o}_{3}$    & 330 506  & 1.0042   \\
14  & $7s7d$            & $^3$D$_{1}$    & 331 078  & 0.5193   \\
15  & $7s8s$            & $^1$S$_{0}$    & 331 644  & 0.0000   \\
16  & $7s7d$            & $^3$D$_{2}$    & 332 414  & 1.1530   \\
17  & $7s7d$            & $^3$D$_{3}$    & 340 445  & 1.3333   \\
18  & $7s7d$            & $^1$D$_{2}$    & 345 117  & 1.0729   \\
19  & $7s8p$            & $^1$S$^{\rm o}_{0}$    & 365 672  & 0.0000   \\
20  & $7s8p$            & $^3$P$^{\rm o}_{1}$    & 366 252  & 1.3436   \\
\hline \hline
    \end{tabular}
\end{table}

\begin{table}[htpb]
    \caption{Ts VII}
    \label{Ts VII}
    \centering
    %
\end{table}

\FloatBarrier

\end{document}

%% file: LvTsSub.bbl
%

%% file: LvTsSub.bbl
\begin{thebibliography}{27}%
\makeatletter
\providecommand \@ifxundefined [1]{%
 \@ifx{#1\undefined}
}%
\providecommand \@ifnum [1]{%
 \ifnum #1\expandafter \@firstoftwo
 \else \expandafter \@secondoftwo
 \fi
}%
\providecommand \@ifx [1]{%
 \ifx #1\expandafter \@firstoftwo
 \else \expandafter \@secondoftwo
 \fi
}%
\providecommand \natexlab [1]{#1}%
\providecommand \enquote  [1]{``#1''}%
\providecommand \bibnamefont  [1]{#1}%
\providecommand \bibfnamefont [1]{#1}%
\providecommand \citenamefont [1]{#1}%
\providecommand \href@noop [0]{\@secondoftwo}%
\providecommand \href [0]{\begingroup \@sanitize@url \@href}%
\providecommand \@href[1]{\@@startlink{#1}\@@href}%
\providecommand \@@href[1]{\endgroup#1\@@endlink}%
\providecommand \@sanitize@url [0]{\catcode `\\12\catcode `\$12\catcode
  `\&12\catcode `\#12\catcode `\^12\catcode `\_12\catcode `\%12\relax}%
\providecommand \@@startlink[1]{}%
\providecommand \@@endlink[0]{}%
\providecommand \url  [0]{\begingroup\@sanitize@url \@url }%
\providecommand \@url [1]{\endgroup\@href {#1}{\urlprefix }}%
\providecommand \urlprefix  [0]{URL }%
\providecommand \Eprint [0]{\href }%
\providecommand \doibase [0]{https://doi.org/}%
\providecommand \selectlanguage [0]{\@gobble}%
\providecommand \bibinfo  [0]{\@secondoftwo}%
\providecommand \bibfield  [0]{\@secondoftwo}%
\providecommand \translation [1]{[#1]}%
\providecommand \BibitemOpen [0]{}%
\providecommand \bibitemStop [0]{}%
\providecommand \bibitemNoStop [0]{.\EOS\space}%
\providecommand \EOS [0]{\spacefactor3000\relax}%
\providecommand \BibitemShut  [1]{\csname bibitem#1\endcsname}%
\let\auto@bib@innerbib\@empty
\bibitem [{\citenamefont {Smits}\ \emph {et~al.}(2023)\citenamefont {Smits},
  \citenamefont {Indelicato}, \citenamefont {Nazarewicz}, \citenamefont
  {Piibeleht},\ and\ \citenamefont {Schwerdtfeger}}]{smits2023}%
  \BibitemOpen
  \bibfield  {author} {\bibinfo {author} {\bibfnamefont {O.}~\bibnamefont
  {Smits}}, \bibinfo {author} {\bibfnamefont {P.}~\bibnamefont {Indelicato}},
  \bibinfo {author} {\bibfnamefont {W.}~\bibnamefont {Nazarewicz}}, \bibinfo
  {author} {\bibfnamefont {M.}~\bibnamefont {Piibeleht}},\ and\ \bibinfo
  {author} {\bibfnamefont {P.}~\bibnamefont {Schwerdtfeger}},\ }\href
  {https://doi.org/https://doi.org/10.1016/j.physrep.2023.09.004} {\bibfield
  {journal} {\bibinfo  {journal} {Physics Reports}\ }\textbf {\bibinfo {volume}
  {1035}},\ \bibinfo {pages} {1} (\bibinfo {year} {2023})}\BibitemShut
  {NoStop}%
\bibitem [{\citenamefont {Smits}\ \emph {et~al.}(2024)\citenamefont {Smits},
  \citenamefont {D{\"u}llmann}, \citenamefont {Indelicato}, \citenamefont
  {Nazarewicz},\ and\ \citenamefont {Schwerdtfeger}}]{smits2024}%
  \BibitemOpen
  \bibfield  {author} {\bibinfo {author} {\bibfnamefont {O.~R.}\ \bibnamefont
  {Smits}}, \bibinfo {author} {\bibfnamefont {C.~E.}\ \bibnamefont
  {D{\"u}llmann}}, \bibinfo {author} {\bibfnamefont {P.}~\bibnamefont
  {Indelicato}}, \bibinfo {author} {\bibfnamefont {W.}~\bibnamefont
  {Nazarewicz}},\ and\ \bibinfo {author} {\bibfnamefont {P.}~\bibnamefont
  {Schwerdtfeger}},\ }\href {https://doi.org/10.1038/s42254-023-00668-y}
  {\bibfield  {journal} {\bibinfo  {journal} {Nature Reviews Physics}\ }\textbf
  {\bibinfo {volume} {6}},\ \bibinfo {pages} {86} (\bibinfo {year}
  {2024})}\BibitemShut {NoStop}%
\bibitem [{\citenamefont {Ackermann}\ \emph {et~al.}(2024)\citenamefont
  {Ackermann}, \citenamefont {Antalic},\ and\ \citenamefont
  {He{\ss}berger}}]{ackermann2024}%
  \BibitemOpen
  \bibfield  {author} {\bibinfo {author} {\bibfnamefont {D.}~\bibnamefont
  {Ackermann}}, \bibinfo {author} {\bibfnamefont {S.}~\bibnamefont {Antalic}},\
  and\ \bibinfo {author} {\bibfnamefont {F.~P.}\ \bibnamefont
  {He{\ss}berger}},\ }\href {https://doi.org/10.1140/epjs/s11734-024-01150-1}
  {\bibfield  {journal} {\bibinfo  {journal} {The European Physical Journal
  Special Topics}\ }\textbf {\bibinfo {volume} {233}},\ \bibinfo {pages} {1017}
  (\bibinfo {year} {2024})}\BibitemShut {NoStop}%
\bibitem [{\citenamefont {Ye}\ \emph {et~al.}(2025)\citenamefont {Ye},
  \citenamefont {Yang}, \citenamefont {Sakurai},\ and\ \citenamefont
  {Hu}}]{ye2025}%
  \BibitemOpen
  \bibfield  {author} {\bibinfo {author} {\bibfnamefont {Y.}~\bibnamefont
  {Ye}}, \bibinfo {author} {\bibfnamefont {X.}~\bibnamefont {Yang}}, \bibinfo
  {author} {\bibfnamefont {H.}~\bibnamefont {Sakurai}},\ and\ \bibinfo {author}
  {\bibfnamefont {B.}~\bibnamefont {Hu}},\ }\href
  {https://doi.org/10.1038/s42254-024-00782-5} {\bibfield  {journal} {\bibinfo
  {journal} {Nature Reviews Physics}\ }\textbf {\bibinfo {volume} {7}},\
  \bibinfo {pages} {21} (\bibinfo {year} {2025})}\BibitemShut {NoStop}%
\bibitem [{\citenamefont {Dzuba}\ \emph
  {et~al.}(2017{\natexlab{a}})\citenamefont {Dzuba}, \citenamefont {Flambaum},\
  and\ \citenamefont {Webb}}]{DFW17-IS}%
  \BibitemOpen
  \bibfield  {author} {\bibinfo {author} {\bibfnamefont {V.~A.}\ \bibnamefont
  {Dzuba}}, \bibinfo {author} {\bibfnamefont {V.~V.}\ \bibnamefont
  {Flambaum}},\ and\ \bibinfo {author} {\bibfnamefont {J.~K.}\ \bibnamefont
  {Webb}},\ }\href@noop {} {\bibfield  {journal} {\bibinfo  {journal} {Phys.
  Rev. A}\ }\textbf {\bibinfo {volume} {95}},\ \bibinfo {pages} {062515}
  (\bibinfo {year} {2017}{\natexlab{a}})}\BibitemShut {NoStop}%
\bibitem [{\citenamefont {Dzuba}\ \emph {et~al.}(2025)\citenamefont {Dzuba},
  \citenamefont {Flambaum},\ and\ \citenamefont {Vong}}]{LvTsNeutral}%
  \BibitemOpen
  \bibfield  {author} {\bibinfo {author} {\bibfnamefont {V.~A.}\ \bibnamefont
  {Dzuba}}, \bibinfo {author} {\bibfnamefont {V.~V.}\ \bibnamefont
  {Flambaum}},\ and\ \bibinfo {author} {\bibfnamefont {G.~K.}\ \bibnamefont
  {Vong}},\ }\href {https://arxiv.org/abs/2505.22895} {\  (\bibinfo {year}
  {2025})},\ \Eprint {https://arxiv.org/abs/2505.22895} {arXiv:2505.22895
  [physics.atom-ph]} \BibitemShut {NoStop}%
\bibitem [{\citenamefont {Safronova}\ \emph {et~al.}(2009)\citenamefont
  {Safronova}, \citenamefont {Kozlov}, \citenamefont {Johnson},\ and\
  \citenamefont {Jiang}}]{Saf-CC}%
  \BibitemOpen
  \bibfield  {author} {\bibinfo {author} {\bibfnamefont {M.~S.}\ \bibnamefont
  {Safronova}}, \bibinfo {author} {\bibfnamefont {M.~G.}\ \bibnamefont
  {Kozlov}}, \bibinfo {author} {\bibfnamefont {W.~R.}\ \bibnamefont
  {Johnson}},\ and\ \bibinfo {author} {\bibfnamefont {D.}~\bibnamefont
  {Jiang}},\ }\href@noop {} {\bibfield  {journal} {\bibinfo  {journal} {Phys.
  Rev. A}\ }\textbf {\bibinfo {volume} {80}},\ \bibinfo {pages} {012516}
  (\bibinfo {year} {2009})}\BibitemShut {NoStop}%
\bibitem [{\citenamefont {Dzuba}(2014)}]{Dzu-CI-SD14}%
  \BibitemOpen
  \bibfield  {author} {\bibinfo {author} {\bibfnamefont {V.~A.}\ \bibnamefont
  {Dzuba}},\ }\href@noop {} {\bibfield  {journal} {\bibinfo  {journal} {Phys.
  Rev. A}\ }\textbf {\bibinfo {volume} {90}},\ \bibinfo {pages} {012517}
  (\bibinfo {year} {2014})}\BibitemShut {NoStop}%
\bibitem [{\citenamefont {Dzuba}\ \emph
  {et~al.}(2017{\natexlab{b}})\citenamefont {Dzuba}, \citenamefont {Berengut},
  \citenamefont {Harabati},\ and\ \citenamefont {Flambaum}}]{CIPT}%
  \BibitemOpen
  \bibfield  {author} {\bibinfo {author} {\bibfnamefont {V.~A.}\ \bibnamefont
  {Dzuba}}, \bibinfo {author} {\bibfnamefont {J.~C.}\ \bibnamefont {Berengut}},
  \bibinfo {author} {\bibfnamefont {C.}~\bibnamefont {Harabati}},\ and\
  \bibinfo {author} {\bibfnamefont {V.~V.}\ \bibnamefont {Flambaum}},\
  }\href@noop {} {\bibfield  {journal} {\bibinfo  {journal} {Phys. Rev. A}\
  }\textbf {\bibinfo {volume} {95}},\ \bibinfo {pages} {012503} (\bibinfo
  {year} {2017}{\natexlab{b}})}\BibitemShut {NoStop}%
\bibitem [{\citenamefont {Kramida}\ \emph {et~al.}()\citenamefont {Kramida},
  \citenamefont {{Yu.~Ralchenko}}, \citenamefont {Reader},\ and\ \citenamefont
  {{NIST ASD Team}}}]{NIST_ASD}%
  \BibitemOpen
  \bibfield  {author} {\bibinfo {author} {\bibfnamefont {A.}~\bibnamefont
  {Kramida}}, \bibinfo {author} {\bibnamefont {{Yu.~Ralchenko}}}, \bibinfo
  {author} {\bibfnamefont {J.}~\bibnamefont {Reader}},\ and\ \bibinfo {author}
  {\bibnamefont {{NIST ASD Team}}},\ }\href@noop {} {\bibinfo {title} {{NIST
  Atomic Spectra Database \textnormal{(ver. 5.12), [Online]. Available: {\tt
  \url{https://physics.nist.gov/asd}} [2024, December 3]. National Institute of
  Standards and Technology, Gaithersburg, MD.}}}},\ \bibinfo {note} {dOI:
  \href{https://doi.org/10.18434/T4W30F}{\tt https://doi.org/10.18434/T4W30F}.
  (2024).}\BibitemShut {Stop}%
\bibitem [{\citenamefont {Dzuba}(2005)}]{Dzu05}%
  \BibitemOpen
  \bibfield  {author} {\bibinfo {author} {\bibfnamefont {V.~A.}\ \bibnamefont
  {Dzuba}},\ }\href@noop {} {\bibfield  {journal} {\bibinfo  {journal} {Phys.
  Rev. A}\ }\textbf {\bibinfo {volume} {71}},\ \bibinfo {pages} {032512}
  (\bibinfo {year} {2005})}\BibitemShut {NoStop}%
\bibitem [{\citenamefont {Johnson}\ and\ \citenamefont
  {Sapirstein}(1986)}]{B-splines}%
  \BibitemOpen
  \bibfield  {author} {\bibinfo {author} {\bibfnamefont {W.~R.}\ \bibnamefont
  {Johnson}}\ and\ \bibinfo {author} {\bibfnamefont {J.}~\bibnamefont
  {Sapirstein}},\ }\href@noop {} {\bibfield  {journal} {\bibinfo  {journal}
  {Phys. Rev. Lett.}\ }\textbf {\bibinfo {volume} {57}},\ \bibinfo {pages}
  {1126} (\bibinfo {year} {1986})}\BibitemShut {NoStop}%
\bibitem [{\citenamefont {Flambaum}\ and\ \citenamefont
  {Ginges}(2005)}]{radpot}%
  \BibitemOpen
  \bibfield  {author} {\bibinfo {author} {\bibfnamefont {V.~V.}\ \bibnamefont
  {Flambaum}}\ and\ \bibinfo {author} {\bibfnamefont {J.~S.~M.}\ \bibnamefont
  {Ginges}},\ }\href@noop {} {\bibfield  {journal} {\bibinfo  {journal} {Phys.
  Rev. A}\ }\textbf {\bibinfo {volume} {72}},\ \bibinfo {pages} {052115}
  (\bibinfo {year} {2005})}\BibitemShut {NoStop}%
\bibitem [{\citenamefont {Dzuba}\ \emph {et~al.}(2001)\citenamefont {Dzuba},
  \citenamefont {Harabati}, \citenamefont {Johnson},\ and\ \citenamefont
  {Safronova}}]{DzuEtAl01a}%
  \BibitemOpen
  \bibfield  {author} {\bibinfo {author} {\bibfnamefont {V.~A.}\ \bibnamefont
  {Dzuba}}, \bibinfo {author} {\bibfnamefont {C.}~\bibnamefont {Harabati}},
  \bibinfo {author} {\bibfnamefont {W.~R.}\ \bibnamefont {Johnson}},\ and\
  \bibinfo {author} {\bibfnamefont {M.~S.}\ \bibnamefont {Safronova}},\
  }\href@noop {} {\bibfield  {journal} {\bibinfo  {journal} {Phys. Rev. A}\
  }\textbf {\bibinfo {volume} {63}},\ \bibinfo {pages} {044103} (\bibinfo
  {year} {2001})}\BibitemShut {NoStop}%
\bibitem [{\citenamefont {Dzuba}\ \emph {et~al.}(2006)\citenamefont {Dzuba},
  \citenamefont {Flambaum},\ and\ \citenamefont {Safronova}}]{DzuFlaSaf06}%
  \BibitemOpen
  \bibfield  {author} {\bibinfo {author} {\bibfnamefont {V.~A.}\ \bibnamefont
  {Dzuba}}, \bibinfo {author} {\bibfnamefont {V.~V.}\ \bibnamefont
  {Flambaum}},\ and\ \bibinfo {author} {\bibfnamefont {M.~S.}\ \bibnamefont
  {Safronova}},\ }\href@noop {} {\bibfield  {journal} {\bibinfo  {journal}
  {Phys. Rev. A}\ }\textbf {\bibinfo {volume} {73}},\ \bibinfo {pages} {022112}
  (\bibinfo {year} {2006})}\BibitemShut {NoStop}%
\bibitem [{\citenamefont {Dzuba}\ \emph {et~al.}(2024)\citenamefont {Dzuba},
  \citenamefont {Flambaum},\ and\ \citenamefont {Afanasjev}}]{Etotal}%
  \BibitemOpen
  \bibfield  {author} {\bibinfo {author} {\bibfnamefont {V.~A.}\ \bibnamefont
  {Dzuba}}, \bibinfo {author} {\bibfnamefont {V.~V.}\ \bibnamefont
  {Flambaum}},\ and\ \bibinfo {author} {\bibfnamefont {A.~V.}\ \bibnamefont
  {Afanasjev}},\ }\href@noop {} {\bibfield  {journal} {\bibinfo  {journal}
  {Phys. Rev. A}\ }\textbf {\bibinfo {volume} {110}},\ \bibinfo {pages}
  {052810} (\bibinfo {year} {2024})}\BibitemShut {NoStop}%
\bibitem [{\citenamefont {Hangele}\ \emph {et~al.}(2012)\citenamefont
  {Hangele}, \citenamefont {Dolg}, \citenamefont {Hanrath}, \citenamefont
  {Cao},\ and\ \citenamefont {Schwerdtfeger}}]{Hangele}%
  \BibitemOpen
  \bibfield  {author} {\bibinfo {author} {\bibfnamefont {T.}~\bibnamefont
  {Hangele}}, \bibinfo {author} {\bibfnamefont {M.}~\bibnamefont {Dolg}},
  \bibinfo {author} {\bibfnamefont {M.}~\bibnamefont {Hanrath}}, \bibinfo
  {author} {\bibfnamefont {X.}~\bibnamefont {Cao}},\ and\ \bibinfo {author}
  {\bibfnamefont {P.}~\bibnamefont {Schwerdtfeger}},\ }\href
  {https://doi.org/10.1063/1.4723805} {\bibfield  {journal} {\bibinfo
  {journal} {The Journal of Chemical Physics}\ }\textbf {\bibinfo {volume}
  {136}},\ \bibinfo {pages} {214105} (\bibinfo {year} {2012})},\ \Eprint
  {https://arxiv.org/abs/https://pubs.aip.org/aip/jcp/article-pdf/doi/10.1063/1.4723805/14097986/214105\_1\_online.pdf}
  {https://pubs.aip.org/aip/jcp/article-pdf/doi/10.1063/1.4723805/14097986/214105\_1\_online.pdf}
  \BibitemShut {NoStop}%
\bibitem [{\citenamefont {Borschevsky}\ \emph {et~al.}(2015)\citenamefont
  {Borschevsky}, \citenamefont {Pa{\v{s}}teka}, \citenamefont {Pershina},
  \citenamefont {Eliav},\ and\ \citenamefont {Kaldor}}]{Borschevsky_2015}%
  \BibitemOpen
  \bibfield  {author} {\bibinfo {author} {\bibfnamefont {A.}~\bibnamefont
  {Borschevsky}}, \bibinfo {author} {\bibfnamefont {L.~F.}\ \bibnamefont
  {Pa{\v{s}}teka}}, \bibinfo {author} {\bibfnamefont {V.}~\bibnamefont
  {Pershina}}, \bibinfo {author} {\bibfnamefont {E.}~\bibnamefont {Eliav}},\
  and\ \bibinfo {author} {\bibfnamefont {U.}~\bibnamefont {Kaldor}},\ }\href
  {https://doi.org/10.1103/PhysRevA.91.020501} {\bibfield  {journal} {\bibinfo
  {journal} {Physical Review A}\ }\textbf {\bibinfo {volume} {91}},\ \bibinfo
  {pages} {020501} (\bibinfo {year} {2015})}\BibitemShut {NoStop}%
\bibitem [{\citenamefont {Dyall}(2012)}]{Dyall2012}%
  \BibitemOpen
  \bibfield  {author} {\bibinfo {author} {\bibfnamefont {K.~G.}\ \bibnamefont
  {Dyall}},\ }\href {https://doi.org/10.1007/s00214-012-1172-4} {\bibfield
  {journal} {\bibinfo  {journal} {Theoretical Chemistry Accounts}\ }\textbf
  {\bibinfo {volume} {131}},\ \bibinfo {pages} {1172} (\bibinfo {year}
  {2012})}\BibitemShut {NoStop}%
\bibitem [{\citenamefont {Liu}\ \emph {et~al.}(2020)\citenamefont {Liu},
  \citenamefont {Shen}, \citenamefont {Wang},\ and\ \citenamefont
  {Sang}}]{Z116}%
  \BibitemOpen
  \bibfield  {author} {\bibinfo {author} {\bibfnamefont {J.}~\bibnamefont
  {Liu}}, \bibinfo {author} {\bibfnamefont {X.}~\bibnamefont {Shen}}, \bibinfo
  {author} {\bibfnamefont {K.}~\bibnamefont {Wang}},\ and\ \bibinfo {author}
  {\bibfnamefont {C.}~\bibnamefont {Sang}},\ }\href
  {https://doi.org/10.1063/5.0007145} {\bibfield  {journal} {\bibinfo
  {journal} {The Journal of Chemical Physics}\ }\textbf {\bibinfo {volume}
  {152}},\ \bibinfo {pages} {204303} (\bibinfo {year} {2020})}\BibitemShut
  {NoStop}%
\bibitem [{\citenamefont {Fricke}\ and\ \citenamefont
  {McMinn}(1976)}]{Fricke1976}%
  \BibitemOpen
  \bibfield  {author} {\bibinfo {author} {\bibfnamefont {B.}~\bibnamefont
  {Fricke}}\ and\ \bibinfo {author} {\bibfnamefont {J.}~\bibnamefont
  {McMinn}},\ }\href {https://doi.org/10.1007/BF00624214} {\bibfield  {journal}
  {\bibinfo  {journal} {Naturwissenschaften}\ }\textbf {\bibinfo {volume}
  {63}},\ \bibinfo {pages} {162} (\bibinfo {year} {1976})}\BibitemShut
  {NoStop}%
\bibitem [{\citenamefont {Liu}\ and\ \citenamefont {Peng}(2006)}]{Liu_2006}%
  \BibitemOpen
  \bibfield  {author} {\bibinfo {author} {\bibfnamefont {W.}~\bibnamefont
  {Liu}}\ and\ \bibinfo {author} {\bibfnamefont {D.}~\bibnamefont {Peng}},\
  }\href {https://doi.org/10.1063/1.2222365} {\bibfield  {journal} {\bibinfo
  {journal} {The Journal of Chemical Physics}\ }\textbf {\bibinfo {volume}
  {125}},\ \bibinfo {pages} {044102} (\bibinfo {year} {2006})}\BibitemShut
  {NoStop}%
\bibitem [{\citenamefont {Chang}\ \emph {et~al.}(2010)\citenamefont {Chang},
  \citenamefont {Li},\ and\ \citenamefont {Dong}}]{Z117}%
  \BibitemOpen
  \bibfield  {author} {\bibinfo {author} {\bibfnamefont {Z.}~\bibnamefont
  {Chang}}, \bibinfo {author} {\bibfnamefont {J.}~\bibnamefont {Li}},\ and\
  \bibinfo {author} {\bibfnamefont {C.}~\bibnamefont {Dong}},\ }\href
  {https://doi.org/10.1021/jp107411s} {\bibfield  {journal} {\bibinfo
  {journal} {The Journal of Physical Chemistry A}\ }\textbf {\bibinfo {volume}
  {114}},\ \bibinfo {pages} {13388} (\bibinfo {year} {2010})},\ \Eprint
  {https://arxiv.org/abs/https://doi.org/10.1021/jp107411s}
  {https://doi.org/10.1021/jp107411s} \BibitemShut {NoStop}%
\bibitem [{\citenamefont {Mitin}\ and\ \citenamefont {van
  Wüllen}(2006)}]{Mitin}%
  \BibitemOpen
  \bibfield  {author} {\bibinfo {author} {\bibfnamefont {A.~V.}\ \bibnamefont
  {Mitin}}\ and\ \bibinfo {author} {\bibfnamefont {C.}~\bibnamefont {van
  Wüllen}},\ }\href {https://doi.org/10.1063/1.2165175} {\bibfield  {journal}
  {\bibinfo  {journal} {The Journal of Chemical Physics}\ }\textbf {\bibinfo
  {volume} {124}},\ \bibinfo {pages} {064305} (\bibinfo {year}
  {2006})}\BibitemShut {NoStop}%
\bibitem [{\citenamefont {Pyper}(2020)}]{Pyper}%
  \BibitemOpen
  \bibfield  {author} {\bibinfo {author} {\bibfnamefont {N.~C.}\ \bibnamefont
  {Pyper}},\ }\href {https://doi.org/10.1098/rsta.2019.0305} {\bibfield
  {journal} {\bibinfo  {journal} {Philosophical Transactions of the Royal
  Society A: Mathematical, Physical and Engineering Sciences}\ }\textbf
  {\bibinfo {volume} {378}},\ \bibinfo {pages} {20190305} (\bibinfo {year}
  {2020})},\ \Eprint
  {https://arxiv.org/abs/https://royalsocietypublishing.org/doi/pdf/10.1098/rsta.2019.0305}
  {https://royalsocietypublishing.org/doi/pdf/10.1098/rsta.2019.0305}
  \BibitemShut {NoStop}%
\bibitem [{\citenamefont {Rodrigues}\ \emph {et~al.}(2004)\citenamefont
  {Rodrigues}, \citenamefont {Indelicato}, \citenamefont {Santos},
  \citenamefont {Patté},\ and\ \citenamefont {Parente}}]{RODRIGUES2004}%
  \BibitemOpen
  \bibfield  {author} {\bibinfo {author} {\bibfnamefont {G.}~\bibnamefont
  {Rodrigues}}, \bibinfo {author} {\bibfnamefont {P.}~\bibnamefont
  {Indelicato}}, \bibinfo {author} {\bibfnamefont {J.}~\bibnamefont {Santos}},
  \bibinfo {author} {\bibfnamefont {P.}~\bibnamefont {Patté}},\ and\ \bibinfo
  {author} {\bibfnamefont {F.}~\bibnamefont {Parente}},\ }\href
  {https://doi.org/https://doi.org/10.1016/j.adt.2003.11.005} {\bibfield
  {journal} {\bibinfo  {journal} {Atomic Data and Nuclear Data Tables}\
  }\textbf {\bibinfo {volume} {86}},\ \bibinfo {pages} {117} (\bibinfo {year}
  {2004})}\BibitemShut {NoStop}%
\bibitem [{\citenamefont {Kaufman}\ \emph {et~al.}(1988)\citenamefont
  {Kaufman}, \citenamefont {Sugar},\ and\ \citenamefont {Joshi}}]{Kaufman}%
  \BibitemOpen
  \bibfield  {author} {\bibinfo {author} {\bibfnamefont {V.}~\bibnamefont
  {Kaufman}}, \bibinfo {author} {\bibfnamefont {J.}~\bibnamefont {Sugar}},\
  and\ \bibinfo {author} {\bibfnamefont {Y.~N.}\ \bibnamefont {Joshi}},\ }\href
  {https://doi.org/10.1364/JOSAB.5.000619} {\bibfield  {journal} {\bibinfo
  {journal} {J. Opt. Soc. Am. B}\ }\textbf {\bibinfo {volume} {5}},\ \bibinfo
  {pages} {619} (\bibinfo {year} {1988})}\BibitemShut {NoStop}%
\end{thebibliography}
